\documentclass[aps,prd,reprint,preprintnumbers,nofootinbib,amsmath,amssymb,longbibliography]{revtex4-2}
\usepackage[utf8]{inputenc}
\usepackage[dvipsnames]{xcolor}
\usepackage{tikz}
\usetikzlibrary{arrows.meta,positioning,bending}
\usepackage[export]{adjustbox}
\usepackage{mathtools,resizegather,amsmath,amsfonts,amssymb,mathrsfs,amsthm,amsbsy,multirow,dcolumn,makecell,scalerel,verbatim,braket,slashed, graphicx, physics,bm, hyperref, upgreek, xfrac}

\usepackage{amsthm}
\usepackage{orcidlink}
\providecommand{\bibinfo}[2]{#2}
\let\originalbibinfo\bibinfo
\renewcommand{\bibinfo}[2]{%
  \ifstrequal{#1}{title}
    {\emph{#2}}
    {\originalbibinfo{#1}{#2}}%
}

\def\be{\begin{equation}}
\def\ee{\end{equation}}

\begin{document}

\preprint{\texttt{IFT-UAM/CSIC-26-163}}

\title{Emergent de Sitter Universes in Non-Hermitian Holography}

\author{Daniel Are\'an,\raisebox{2.5pt}{\scalebox{0.80}{\orcidlink{0000-0003-2164-2777}}}$^{\,1,2}$ Suman Das,\raisebox{2.5pt}{\scalebox{0.80}{\orcidlink{0000-0002-0053-3187}}}$^{\,1}$ David Garcia-Fari\~na,\raisebox{2.5pt}{\scalebox{0.80}{\orcidlink{0009-0008-3058-7371}}}$^{\,1,2}$ Juan F. Pedraza\raisebox{2.5pt}{\scalebox{0.80}{\orcidlink{0000-0002-1426-642X}}}$^{\,1}$}
\affiliation{$^{1}$Instituto de F\'isica Te\'orica UAM/CSIC, 28049 Madrid, Spain\\$^2$Departamento de F\'isica Te\'orica, Universidad Aut{\'o}noma de Madrid, 28049 Madrid, Spain}

\begin{abstract}
We construct a class of asymptotically anti-de Sitter (AdS), regular black holes whose interiors evolve into nonsingular, expanding de Sitter (dS) cosmologies. Our construction relies on non-Hermitian holography, where violations of the null energy condition allow the interior geometry to undergo a smooth bounce and approach a late-time dS attractor. The cosmology emerges dynamically behind a single event horizon, extending to a spacelike future boundary, $\mathcal I^+$, without introducing auxiliary branes or gluing distinct geometries. This provides a novel setting to explore the connections between three complementary holographic perspectives: a non-Hermitian conformal field theory (CFT) at the AdS boundaries, a putative Euclidean CFT at $\mathcal I^+$ (dS/CFT), and observer-centered static-patch holography. Remarkably, observables anchored at the AdS boundaries reveal concrete connections with these complementary notions of dS holography. Our construction thus offers a new holographic laboratory for systematically exploring how future-boundary and static-patch descriptions of dS quantum gravity may emerge within the established setting of AdS/CFT.
\end{abstract}

\maketitle
\flushbottom

%%=============================%%
\textbf{Introduction.}
%%=============================%%
Can an asymptotically AdS black hole contain a portion of a dS universe? Such a possibility would uncover a striking connection between two very different settings: AdS space, where holography provides a well-established
framework for quantum gravity, and dS space, whose
microscopic description remains elusive. Realizing this is nevertheless challenging,
as the focusing constraints imposed by the null energy
condition (NEC) obstruct the nonsingular transition from a contracting black-hole interior to an expanding cosmology \cite{Farhi:1986ty}.

Several approaches have pursued this idea. Early constructions
considered inflating dS bubbles embedded in asymptotically
AdS spacetimes \cite{Freivogel:2005qh}, while centaur geometries exploit the special features of
two-dimensional dilaton gravity to circumvent the usual higher-dimensional focusing constraints, interpolating between AdS$_2$ and dS$_2$ static patches
\cite{Anninos:2017hhn,Anninos:2018svg}. Braneworld holography has also provided explicit realizations of dS geometries within higher-dimensional AdS spacetimes \cite{Emparan:2022ijy,Panella:2023lsi,Climent:2024wol,Panella:2024sor,Aguilar-Gutierrez:2023tic}, however, the dS geometry is supported on auxiliary branes, rather than emerging dynamically within the higher-dimensional bulk. Despite this progress, realizing an expanding dS cosmology directly inside a higher-dimensional AdS black hole, without introducing auxiliary branes or gluing distinct geometries, remains an important challenge.

The holographic implications of such a construction
are equally compelling. In particular, the dS/CFT correspondence
conjectures that dS quantum gravity admits a description
in terms of a Euclidean CFT at future infinity,
$\mathcal{I}^+$ \cite{Witten:2001kn,Strominger:2001pn},
with the cosmological wavefunction playing a central
role in the proposed dictionary \cite{Maldacena:2002vr,Harlow:2011ke}.
A complementary, observer-centered perspective is
offered by static-patch holography. Proposals
include dS/dS \cite{Alishahiha:2004eh,Alishahiha:2005dj},
involving cutoff CFTs coupled to gravity;
radial cutoff holography
\cite{Gorbenko:2018oov,Coleman:2021nor,Silverstein:2024xnr},
via $T\bar T+\Lambda_2$ and $T^2$ deformations;
and stretched-horizon holography
\cite{Susskind:2021esx,Shaghoulian:2021cef},
which places holographic degrees of freedom
on stretched cosmological horizons.
These developments have motivated studies of
entanglement, complexity, and quantum chaos
\cite{Shaghoulian:2022fop,Jorstad:2022mls,Aguilar-Gutierrez:2023zqm,Ahn:2025qat,Begines:2026fnx,Harlow:2026pwe}.
Yet the microscopic foundations of these descriptions,
and their relation to one another, remain poorly
understood. This raises a natural
question: can dS physics, including aspects of both
its future-boundary and static-patch descriptions, emerge
within a single asymptotically AdS holographic theory?

In this Letter, we provide a concrete setting to
investigate this question. Using $\mathcal{PT}$-symmetric
non-Hermitian holography \cite{Arean:2019pom}, whose
real, $\mathcal{PT}$-restored 
phase permits NEC violations
\cite{Xian:2023zgu,Arean:2026jie,Begines:2026mlv}, we construct asymptotically AdS
black holes with nonsingular, expanding dS interiors.
Previous work showed that NEC violation can replace
the Kasner singularity with a cosmology exhibiting
Little Rip behavior \cite{Arean:2026jie}. Here, a
simple modification of the scalar potential produces
a qualitatively different outcome: the interior
undergoes a smooth bounce and approaches a late-time
dS attractor, with finite curvature and a Hubble
scale determined by the scalar potential.

A distinctive feature of our solutions is their global causal
structure. Unlike many conventional regular black holes,
which possess an inner Cauchy horizon separating the
exterior from a dS-like core
\cite{Hayward:2005gi,Bueno:2024dgm,Bueno:2024eig},
our geometries have a single event horizon. Beyond it,
the interior evolves smoothly toward a spacelike dS
future infinity, $\mathcal{I}^+$. Crucially, the emergent
dS region remains connected to the AdS boundaries
through spacelike bulk slices. This provides a setting
to explore the relation between AdS boundary data and
cosmological states at $\mathcal{I}^+$, building on recent
work on the holographic encoding of black-hole interiors
\cite{Blacker:2026poc}. Meanwhile, the emergent
cosmology admits observer-dependent horizons, opening
a complementary perspective through static-patch holography.

These complementary perspectives motivate us to
investigate how the emergent cosmology manifests
itself in observables of the AdS boundary theory.
We show that two-sided correlators encode asymptotic
data at $\mathcal{I}^+$, while holographic complexity exhibits features characteristic of dS static-patch holography. We also investigate quantum chaos through gravitational shockwaves and
pole-skipping, probing the scrambling dynamics
associated with boundary four-point out-of-time-order
correlators (OTOCs) and contrasting the black-hole response
with that of the cosmological horizon.
Our construction thus opens a new route toward
understanding dS physics, including its future-boundary
and static-patch descriptions, within the established
framework of AdS/CFT.

\vspace{1.5mm}
%%=================================%%
\textbf{Setup.}
%%=================================%%
Our starting point is a
$\mathcal{PT}$-symmetric non-Hermitian QFT in
$2+1$ dimensions \cite{Arean:2019pom},
obtained by deforming a Hermitian
CFT with a charged scalar operator $\mathcal O$
of dimension $\Delta$. Schematically,
\begin{equation}
S_{\rm QFT}=S_{\rm CFT} +\int \dd^3x\,M\left[\left(\mathcal{O}+\bar{\mathcal{O}}\right)+\eta\left(\mathcal{O}- \bar{\mathcal{O}}\right)\right],
\label{eq:CFTdeformation}
\end{equation}
where $\eta$ is a real parameter controlling the
non-Hermitian deformation. Its gravitational dual is described by an
Einstein--Maxwell--scalar theory, with action
\begin{equation}
S=\int \dd^4x\,\sqrt{-g}\left[
R+6-\frac{1}{4}F_{\mu\nu}F^{\mu\nu}
-D_\mu\phi D^\mu\bar\phi-V(\phi\bar\phi)
\right],
\label{eq:action}
\end{equation}
where the AdS radius is set to unity.
The non-Hermiticity is encoded in boundary
conditions for the scalars $\phi$ and $\bar\phi$,
dual to $\mathcal O$ and $\bar{\mathcal O}$,
respectively. Specifically, the deformation
\eqref{eq:CFTdeformation} is implemented by
parameterizing
\begin{equation}
\phi=(1-\eta)\psi,\qquad
\bar\phi=(1+\eta)\psi,
\end{equation}
with $\psi\sim Mz^{3-\Delta}$ at the AdS boundary
$z=0$. We consider neutral ($A_\mu=0$),
homogeneous planar black holes. For $|\eta|>1$,
these solutions violate the bulk NEC
\cite{Xian:2023zgu}. The ansatz, equations of
motion, and numerical construction are detailed
in Appendix~\hyperref[subend:numerics]{A} of the End Matter.

To obtain black holes with dS interiors, we consider
a quartic scalar potential of the form
\begin{equation}
V(\phi\bar\phi)=m^2\phi\bar\phi
-v(\phi\bar\phi)^2,
\label{eq:AdSpotential}
\end{equation}
with $m^2=-2$ ($\Delta=2$).
Here the sign convention for $v$ is opposite to that
of~\cite{Arean:2026jie}, whose Little Rip interiors
are recovered for $v<0$ in our notation.
For $v>0$, the potential admits a nontrivial
stationary point at $\phi\bar\phi=-1/v$, which
supports the emergent dS attractor.

\begin{figure}[t!]
    \centering
    \includegraphics[width=0.99\linewidth]{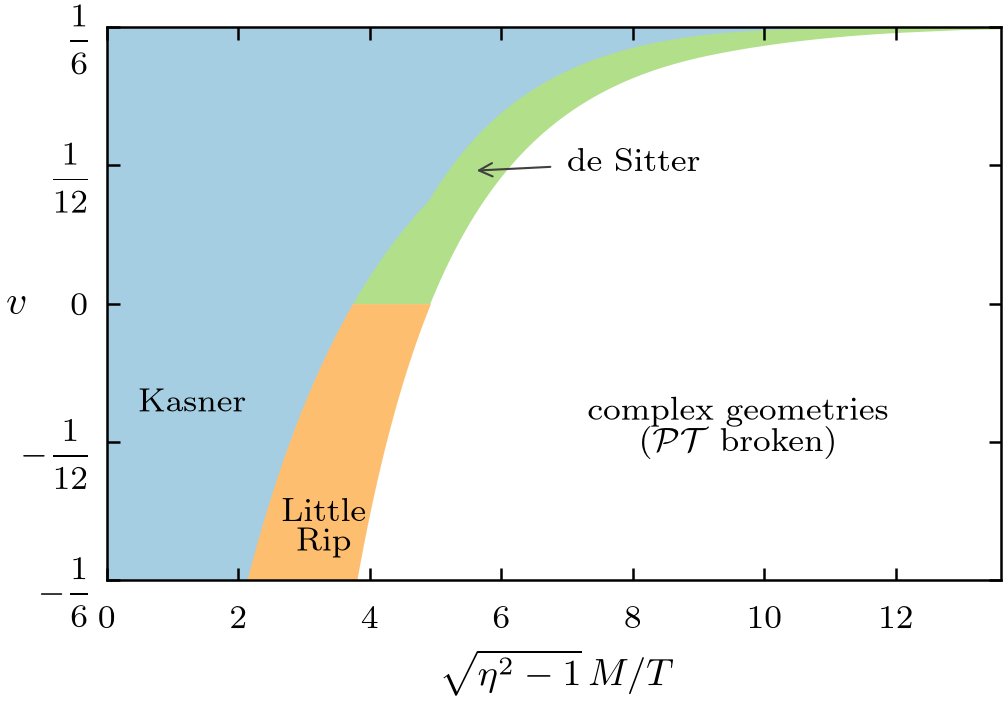}\par
    \vspace{-2mm}
    \caption{Phase diagram of the non-Hermitian model
    as a function of $v$ and $\sqrt{\eta^2-1}\,M/T$.
    The colored regions correspond to Kasner interiors
    (blue), Little Rip cosmologies (orange), and regular
    dS interiors (green). The white region contains
    complex geometries with broken $\mathcal{PT}$
    symmetry.}
    \label{fig:phase_diagram}
\end{figure}

The phase diagram is shown in
Fig.~\ref{fig:phase_diagram}, as a function of $v$
and the dimensionless deformation
$\sqrt{\eta^2-1}\,M/T$. 
Depending on these parameters, the interior
exhibits Kasner-like behavior, Little Rip
evolution, or a nonsingular dS cosmology. 
The dS solutions occupy a finite region 
bounded by Kasner geometries and a domain
of complex solutions where $\mathcal{PT}$ symmetry
is broken. 

\begin{figure}[t!]
    \centering
    \includegraphics[clip, trim=0 8mm 0 5mm, width=0.75\linewidth]{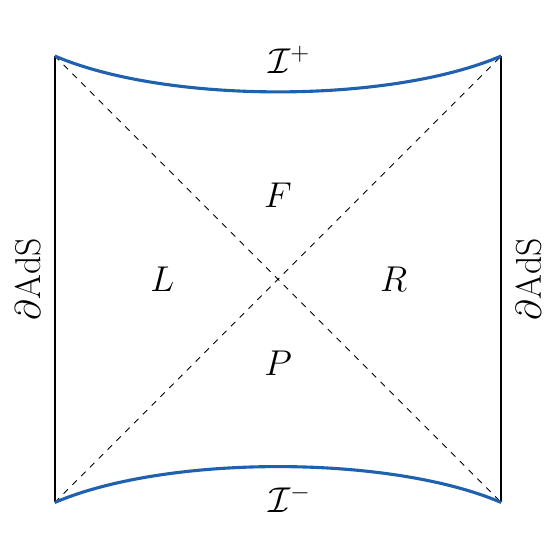}
    \caption{Penrose diagram of an eternal AdS black hole
    with a dS cosmological interior. $L$ and $R$ denote
    the asymptotic AdS boundaries, while $\mathcal I^+$
    and $\mathcal I^-$ are the spacelike future and past
    dS boundaries, respectively. The geometry contains
    a single event horizon and no inner Cauchy horizon.}
    \label{fig:penrose_diagram}
\end{figure}

In the dS case, the interior undergoes
a smooth bounce and approaches an expanding
dS phase.
The geometry interpolates between the timelike AdS conformal
boundary to the spacelike dS future conformal
boundary $\mathcal I^+$,
\begin{align}
ds^2\big|_{z\to0^+}
&\simeq\frac{1}{z^2}
\left(-dt^2+d\vec x^{\,2}+dz^2\right),
\label{eq:AdSmetric}\\
ds^2\big|_{\tau\to0^-}
&\simeq\frac{1}{H^2\tau^2}
\left(c_0^2dt^2+d\vec x^{\,2}-d\tau^2\right),
\label{eq:dsmetric}
\end{align}
where $c_0$ is fixed by the bulk solution
after imposing $g_{tt}=1/z^2$ at the AdS boundary.
The Hubble constant is determined by the potential
at its stationary point,
\begin{equation}
H^2=-1+\frac{m^4}{24v}
=-1+\frac{1}{6v}.
\label{eq:dSHubble}
\end{equation}
A positive dS curvature therefore requires
$0<v<1/6$. The resulting spacetime has a single
event horizon, with the expanding dS region
extending to $\mathcal I^+$.
The corresponding global causal structure is illustrated in
Fig.~\ref{fig:penrose_diagram}.

\vspace{1.5mm}
%%====================================%%
\textbf{The QFT at $\mathcal{I}^+$.}\label{sec:CFTplusPerspective}
%%====================================%%
The eternal black hole of Fig.~\ref{fig:penrose_diagram}
admits multiple holographic perspectives.
In the standard AdS/CFT description, the geometry
is dual to a thermofield double (TFD) state of
QFTs at the two asymptotic AdS boundaries
\cite{Maldacena:2001kr}.
The presence of a spacelike dS future boundary
$\mathcal I^+$ offers a complementary perspective
through dS/CFT. Following
\cite{Strominger:2001pn,Maldacena:2002vr},
the late-time cosmological wavefunction
$\Psi_{\mathcal I^+}$ is identified with the
partition function of a putative Euclidean QFT
living at $\mathcal I^+$,
\begin{equation}
\mathcal Z_{{\rm QFT}^+}
=\Psi_{\mathcal I^+}\approx e^{iS_{\rm ren}},
\label{eq:ZQFTplus}
\end{equation}
which we denote by QFT$^+$ to distinguish it
from the standard theories at the AdS boundaries.

To characterize this Euclidean theory, we expand
around the stationary point
of the scalar potential~\eqref{eq:AdSpotential}, located at
$\phi\bar\phi=m^2/(2v)=-1/v$.
To this end, we express the scalar fields
in St\"uckelberg form,
\begin{equation}
\phi=\rho e^{iq\theta},\qquad
\bar\phi=\rho e^{-iq\theta},
\label{eq:stuckelberg}
\end{equation}
where, unlike in a Hermitian theory, $\rho$ and $\theta$
are generally complex. The dS stationary point corresponds to
$\rho_*^2=-1/v$.
Expanding the scalar field around this extremum
as $\rho=\rho_*+\sigma$ and introducing the
gauge-invariant combination
$B_\mu\equiv\partial_\mu\theta-A_\mu$,
the action~\eqref{eq:action} becomes
\begin{equation}
\begin{aligned}
S=\int \dd^4x\,\sqrt{-g}\Big[
&R-6H^2-\frac14 F_{\mu\nu}F^{\mu\nu}-\partial_\mu\sigma\partial^\mu\sigma\\
&-q^2(\rho_*+\sigma)^2B_\mu B^\mu
-\tilde V(\sigma)\Big],
\end{aligned}
\label{eq:dSaction}
\end{equation}
with $H$ given by Eq.~\eqref{eq:dSHubble} and
\begin{equation}
\tilde V(\sigma)
=-2m^2\sigma^2-4v\rho_*\sigma^3-v\sigma^4.
\label{eq:dsPotential}
\end{equation}
The cubic interaction is imaginary, reflecting
the non-Hermitian nature of the dS extremum.
The scalar $\sigma$ has mass
$m_\sigma^2=-2m^2=4$ in the asymptotic dS
geometry. Near $\mathcal I^+$, its
characteristic modes behave as
\begin{equation}
\sigma\approx
\sigma_s\tau^{\Delta_-}
+\sigma_{\rm v}\tau^{\Delta_+},
\end{equation}
with
\begin{equation}
\Delta_\pm=\frac32\pm
\sqrt{\frac94-\frac{m_\sigma^2}{H^2}}
=\frac32\pm\sqrt{\frac94-\frac4{H^2}}\,.
\label{eq:dSdimension}
\end{equation}
In dS/CFT, $\sigma$
is dual to a scalar operator $\mathcal O_\sigma$
of QFT$^+$ with scaling dimension $\Delta_+$, sourced
by $\sigma_s$, while $\sigma_{\rm v}$ encodes its
response. 
For $v>3/50$, $\Delta_\pm$ become
complex, as expected from the representation
theory of dS. As in AdS/CFT, we interpret QFT$^+$
as a non-Hermitian CFT$^+$ whose conformal
symmetry is explicitly broken by the purely
imaginary source $\sigma_s$. 
Source-free configurations with $\sigma_s=0$
and $\sigma_{\rm v}\neq0$ instead correspond to
spontaneous breaking of Euclidean conformal
symmetry, analogous to vev-driven solutions
in AdS/CFT \cite{Bianchi:2001de,Skenderis:2006di}. 
Unlike its AdS counterpart,
however, CFT$^+$ is intrinsically non-unitary
\cite{Strominger:2001pn,Maldacena:2002vr},
so its non-Hermiticity cannot be attributed
solely to the scalar source.

The gravitational equations provide a further
link between the two descriptions: the
renormalized dS and AdS stress tensors,
$\mathcal T_{\mu\nu}$ and $T_{\mu\nu}$,
respectively, satisfy
\begin{equation}\label{eq:identity}
-ic_0\left(
\expval{\mathcal T^t{}_t}
-\expval{\mathcal T^1{}_1}
\right)
=-\left(
\expval{T^t{}_t}
-\expval{T^1{}_1}
\right)=sT,
\end{equation}
where $s$ and $T$ are the entropy density and
Hawking temperature of the bulk black-hole
horizon. Their product gives the
enthalpy density of the AdS thermal state.
At $\mathcal I^+$, however, $t$ labels a spatial
direction, and this same thermodynamic quantity
is instead encoded in the anisotropy of the Euclidean
stress tensor. For real bulk solutions, this
tensor is purely imaginary, as expected from
the wavefunction prescription~\eqref{eq:ZQFTplus},
so the factor of $-i$ renders the relation real.
In this way, Eq.~\eqref{eq:identity} identifies the
Euclidean stress-tensor anisotropy as a direct
diagnostic of the black-hole enthalpy density
$sT$, linking horizon thermodynamics to
one-point functions at $\mathcal I^+$.
Explicit expressions for both stress tensors
are given in Appendices
\hyperref[subend:AdSHoloRenrom]{B} and
\hyperref[subend:dSHoloRenrom]{C} of the End Matter.

\vspace{1.5mm}
%%====================================%%
\textbf{Holographic signatures.}
\label{sec:observables}
%%====================================%%
We now investigate how the emergent dS cosmology
is encoded in observables of the AdS boundary theory.
We focus on three complementary probes: two-sided
correlators, holographic complexity, and scrambling
dynamics. Detailed calculations are provided in the
Supplemental Material.

\emph{Two-sided correlators.---}
Analytically continued two-sided correlators of heavy operators
provide a window into black-hole interiors. Originally used
to probe Schwarzschild singularities
\cite{Fidkowski:2003nf}, this approach was recently extended to
extract Kasner exponents from nonanalytic corrections
\cite{Frenkel:2020ysx}, and more recently to identify
signatures of Little Rip cosmologies \cite{Arean:2026jie}.
Here, we apply the same strategy to the emergent dS region.

For a heavy probe operator of dimension $\Delta_H\gg1$,
the two-sided correlator in the thermofield double
(TFD) state admits the geodesic approximation
\cite{Balasubramanian:1999zv,Louko:2000tp}, 
\begin{equation}
\left\langle
\mathcal O_H(t-i\beta/2)\mathcal O_H(-t)
\right\rangle
\sim e^{-\Delta_H L_{\rm ren}},
\label{eq:two_sided}
\end{equation}
where $L_{\rm ren}$ is the renormalized length of a spacelike
geodesic connecting the two AdS boundaries. As the geodesic
approaches a null trajectory, it penetrates arbitrarily
deep into the dS region, approaching
$\mathcal I^+$. Remarkably, the near-dS geometry
produces nonanalytic contributions
that encode asymptotic cosmological data.

To illustrate this, we consider $m^2=-2$ and
$v=1/18$, for which the operator $\mathcal{O}_\sigma$ of
QFT$^+$ has scaling dimension $\Delta_+=2$.
In this case, we find
\begin{equation}
L_{\rm ren}\supset \frac{\pi}{H}
-\frac{v\rho_*\sigma_s^3}{4}
(c_0\Delta t)^3\log^2(c_0\Delta t),
\label{eq:dS_signature}
\end{equation}
where $\Delta t=t-t_{\rm null}$, and $\sigma_s$
is the leading coefficient in the asymptotic
expansion of $\sigma$ near $\mathcal I^+$.
The constant term reflects the dS curvature
scale, while the nonanalytic contribution encodes
$\sigma_s$, naturally interpreted as a source
for $\mathcal{O}_\sigma$ in the proposed dS/CFT dictionary.
Accompanying single-logarithmic terms also probe
subleading metric data at $\mathcal I^+$,
potentially related to the cosmological stress tensor.
Thus, the analytic structure of AdS boundary
correlators provides access to quantities
characterizing the emergent cosmological boundary,
offering a concrete starting point for relating
AdS/CFT and dS/CFT.

\emph{Holographic complexity.---}
Holographic complexity offers a complementary probe of
black-hole interiors, with Complexity$\,=\,$Volume (CV)
and Complexity$\,=\,$Action (CA) among its most prominent
proposals \cite{Stanford:2014jda,Brown:2015bva}.
Early studies of two-dimensional centaur geometries
revealed a striking departure from conventional late-time
growth: extremal lengths decrease at a divergent rate
and cease to exist beyond a finite boundary time
\cite{Chapman:2021eyy}.
In higher-dimensional dS static-patch holography,
both CV and CA exhibit the related phenomenon of
\emph{hyperfast growth}, whereby the complexity
growth rate diverges at finite boundary time
\cite{Jorstad:2022mls}. The broader
Complexity$\,=\,$Anything (CAny) framework
\cite{Belin:2021bga}, however, has since shown that this
behavior is not universal: depending on the prescription,
complexity in dS can instead exhibit conventional
linear growth at arbitrarily late times
\cite{Aguilar-Gutierrez:2023zqm,Aguilar-Gutierrez:2023tic}.

CAny has also proved useful for probing singularities of AdS black holes \cite{Jorstad:2023kmq}, including extracting
Kasner exponents from late-time complexity
growth \cite{Arean:2024pzo}. What happens to this
picture when the singularity is replaced by an
expanding dS cosmology?
To address this, we consider CAny proposals
defined on constant mean curvature (CMC) slices
with prescribed curvature $K$. In our geometries,
the homogeneity of the interior ensures that
spatial slices are CMC.
Unlike Kasner geometries, where $K$ diverges near
the singularity, our geometry approaches dS space
with $K\to3H$.
Crucially, we find $K$ stays positive
throughout the interior and has a finite minimum,
\begin{equation}
K_{\rm c}\equiv\min |K|>0.
\label{eq:Kc}
\end{equation}
This critical curvature separates two distinct regimes.
For $|K|>K_{\rm c}$, the CMC slices approach
stationary configurations, yielding linear late-time
complexity growth. For $|K|<K_{\rm c}$,
no stationary final slice exists; instead, the slices
reach $\mathcal I^+$ at finite boundary time, mirroring
the geometric mechanism underlying hyperfast growth.
Remarkably, both regimes emerge from observables
anchored at the AdS boundaries rather than stretched
horizons, revealing a striking connection
with dS static-patch holography. It would be
interesting to understand whether these distinct
behaviors reflect different costs for preparing
states, building on recent progress in relating
bulk observables to boundary generators
\cite{Caceres:2025myu,Carrasco:2026bjx}.

\emph{Quantum chaos.---}
A final probe concerns the scrambling properties of
gravitational horizons. In two-dimensional centaur
geometries, OTOCs exhibit oscillatory rather than
conventional exponential Lyapunov behavior
\cite{Anninos:2018svg}.
Higher-dimensional dS shockwave analyses instead
revealed maximal Lyapunov growth
\cite{Aalsma:2020aib,Aalsma:2021kle},
while localized shockwaves and pole-skipping
in Schwarzschild--de Sitter (SdS) uncovered
unconventional butterfly velocities: superluminal
at sufficiently small black-hole horizons and
imaginary at cosmological horizons
\cite{Ahn:2025qat}.

We first examine the black-hole horizon, which encodes scrambling in the AdS boundary theory.
Independent shockwave and pole-skipping calculations
\cite{Shenker:2013pqa,Roberts:2014isa,Grozdanov:2017ajz,Blake:2017ris}
yield
\begin{equation}
\lambda_L=2\pi T,\qquad
v_B^2=\frac{(2\pi T)^2}{3-V_h/2},
\label{eq:chaos_result}
\end{equation}
where $V_h$ is the value of the potential at the horizon.
Remarkably, the Lyapunov exponent remains maximal
in this non-Hermitian setting, where the
unitarity assumptions underlying the chaos bound
need not hold \cite{Maldacena:2015waa}.
By contrast, NEC violation enhances $v_B$ beyond
the bound $v_B\le\sqrt{3}/2$ for asymptotically
AdS$_4$ planar black branes with NEC-satisfying matter
\cite{Mezei:2016wfz,Mezei:2016zxg}.
For our dS solutions, $v_B$ becomes superluminal
over nearly the entire allowed range of
$\sqrt{\eta^2-1}\,M/T$. This parallels the SdS
black-hole-horizon result \cite{Ahn:2025qat},
where superluminal $v_B$ was linked to possible
nonlocality of the dual theory.
Similar enhancements have also been observed in nonlocal
holographic theories, including noncommutative and
dipole-deformed super Yang--Mills
\cite{Fischler:2018kwt,Eccles:2021zum}.
Here, by contrast, the boundary theory is local but
non-Hermitian, with NEC violation allowing the enhancement.

A complementary diagnostic arises from the
cosmological horizons of observers in the
emergent dS region. Unlike SdS, our interior
is a time-dependent, anisotropic cosmology,
approaching pure dS only asymptotically.
In this late-time limit, a comoving observer acquires
a stationary cosmological horizon with
$T_{\rm c}=H/(2\pi)$. The corresponding
pure-dS shockwave analysis predicts
\begin{equation}
\lambda_{L,\rm c}=2\pi T_c,\qquad
v_{B,\rm c}^2=-\frac12,
\label{eq:cosmo_chaos}
\end{equation}
where $v_{B,\rm c}$ is normalized using physical
distances along the cosmological horizon.
Crucially, these chaos parameters characterize
the asymptotic static patch, whose time evolution
differs from that of the AdS boundary theory.
They therefore cannot be directly identified with
its thermal OTOCs, but may instead be related
to those of dS static-patch observers.
Indeed, recent work identified anti-scrambling
in such correlators \cite{Harlow:2026pwe}.
In this light, the imaginary butterfly velocity,
which signals an oscillatory spatial response
rather than a conventional scrambling front
\cite{Ahn:2025qat}, may be viewed as a
complementary spatial manifestation of
unconventional static-patch dynamics.
Interestingly, such oscillatory behavior
has also been
linked to non-Hermitian dynamics
\cite{Ahn:2025qat}. Since both boundary QFTs
in our setup are non-Hermitian, an effective
static-patch description may inherit this feature.

\vspace{1.5mm}
\textbf{Discussion.}
Our results establish a new mechanism by which
expanding dS cosmologies emerge dynamically within
asymptotically AdS black holes, replacing the
 spacelike singularity with a smooth,
nonsingular future. Crucially, this transition
takes place entirely within a single gravitational
spacetime, without introducing auxiliary branes
or gluing distinct geometries. Beyond
providing a novel realization of regular black
holes, our construction opens two promising
directions toward understanding the microscopic
foundations of dS holography.

First, our setup suggests an unconventional
realization of double holography, with three
complementary descriptions of the same physical
system: a gravitational bulk and two distinct
boundary QFTs. Through AdS/CFT, the bulk
geometry is dual to a non-Hermitian CFT at
the asymptotic AdS boundaries. Meanwhile,
its expanding interior approaches a spacelike
future boundary $\mathcal I^+$, where dS/CFT
suggests a second, putative Euclidean CFT
description.
Our results establish concrete connections between
these perspectives. At the level of one-point
functions, our identity~\eqref{eq:identity} relates
the enthalpy density encoded in the AdS boundary
stress tensor to a spatial anisotropy of the
Euclidean stress tensor at $\mathcal I^+$.
Beyond this, analytically continued two-sided
AdS correlators encode asymptotic scalar data
at $\mathcal I^+$, naturally interpreted as sources
in the Euclidean theory, as shown in
Eq.~\eqref{eq:dS_signature}. Thus, both thermodynamic
data and correlation functions of the AdS
theory carry information about the emergent
cosmological boundary. Establishing a more
complete dictionary relating sources, expectation
values, higher-point correlators, and cosmological
wavefunctions is an important next step.

This framework also opens the possibility of
investigating further observables associated with the
QFT$^+$. Examples include extremal surfaces,
entanglement wedges, pseudo-entropies, R\'enyi
entropies, and complexity, previously explored
in formulations of dS/CFT
\cite{Narayan:2017xca,Narayan:2020nsc,Doi:2022iyj,Nanda:2025tid}.
Of particular interest is a recent proposal
relating Krylov complexity to extremal timelike
volumes anchored at $\mathcal I^\pm$, rather
than stretched cosmological horizons
\cite{Heller:2025ddj}. Comparing such
constructions with our AdS-boundary-anchored
observables could clarify which features
of the emergent cosmology are accessible
in each description and how the two
holographic perspectives are related.

Second, our results suggest that the same geometry
may also admit an observer-centered description
through dS static-patch holography.
This perspective is complementary to dS/CFT,
focusing on the physics accessible within
a cosmological horizon. Depending on
the prescription, AdS-boundary-anchored complexity
exhibits either linear late-time growth or
finite-time behavior reminiscent of hyperfast
growth in dS static patches. Meanwhile,
shockwaves at the asymptotic cosmological
horizon exhibit an imaginary butterfly velocity
and oscillatory spatial response, as found in
\cite{Ahn:2025qat}, echoing the anti-scrambling
behavior of dS static-patch observers
\cite{Harlow:2026pwe}.

These parallels motivate a possible
first-principles derivation of static-patch
holography. Rather than postulating independent
microscopic degrees of freedom on stretched
cosmological horizons, one could seek to derive
their effective description by coarse-graining
the original non-Hermitian AdS boundary theory.
Holographic Wilsonian renormalization provides
a natural framework, relating the integration
of ultraviolet degrees of freedom to the removal
of bulk geometry
\cite{Heemskerk:2010hk,Faulkner:2010jy}.
Here, one could envisage integrating
out the region between the AdS boundaries and
the stretched horizons of the
emergent dS cosmology. The remaining interior
would be encoded in an effective, generally
nonlocal theory inherited from the original
non-Hermitian CFT. The recently proposed
holographic banners, which use the
Hamilton--Jacobi framework to connect AdS
boundary data with interior cosmological states
\cite{Blacker:2026poc}, together with generalized
holographic coarse-graining prescriptions and
their associated induced boundary actions
\cite{Guijosa:2022jdo}, offer complementary
starting points for developing this picture.

\begin{figure}[t!]
    \centering
    \includegraphics[clip, trim=0 0 0 -1mm,width=.99\linewidth]{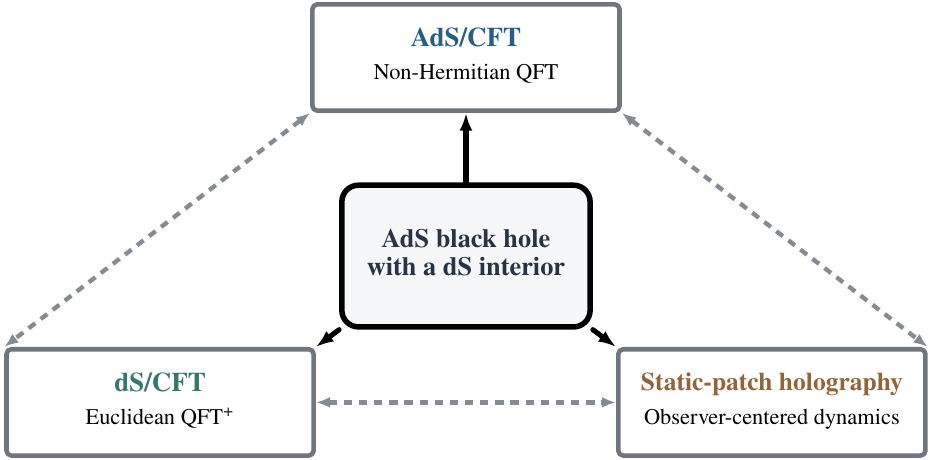}\par
\vspace{-2mm}
\caption{Complementary holographic perspectives within our framework. The same bulk geometry provides a common setting for three distinct holographic descriptions: non-Hermitian AdS/CFT, Euclidean dS/CFT, and observer-centered static-patch holography. Dashed arrows indicate possible connections between these three viewpoints.
}
\label{fig:holo_triangle}
\end{figure}

Taken together, these connections suggest a unified
holographic picture involving three complementary
perspectives, as illustrated in Fig.~\ref{fig:holo_triangle}.
The distinctive feature of our framework is that
the cosmological future emerges dynamically within
a single AdS black hole, rather than being introduced
as an independent ingredient.
Our construction thus provides a concrete setting
to investigate how future-boundary and static-patch
descriptions of dS quantum gravity may arise from
the same non-Hermitian AdS boundary theory and,
ultimately, how their microscopic degrees of freedom
may be related.

\vspace{0.15cm}
\noindent\emph{\textbf{Note added:}} While completing this work, we became aware of
an independent study of closely related
higher-dimensional AdS-to-dS flows supported
by ghost scalars \cite{BeginesAdSdS}.
That study focuses on thermodynamics and
linear stability, and investigates
possible higher-dimensional uplifts.
Our work instead explores the holographic
implications of such geometries, particularly
their potential to connect AdS/CFT with
future-boundary and static-patch descriptions
of dS quantum gravity.

\vspace{0.15cm}
\noindent \emph{Acknowledgements.} 
We thank José Barbón, José Begines, Matthew Blacker, Rafael Carrasco, Antón Faedo, Carlos Hoyos, Karl Landsteiner, Daniele Musso, and Andrew Svesko for interesting discussions and comments on the manuscript. 
This work is supported through the grants CEX2025-001574-S and PID2024-156043NB-I00 funded by MCIN/AEI/10.13039/501100011033 and by ERDF, EU. The work of DGF is supported by FPI grant PRE2022-101810. SD and JFP are supported by the ‘Atracción de Talento’ program of the Comunidad de Madrid under grant 2020-T1/TIC-20495. The research presented in this publication falls within the research line Strings and Quantum Gravity.

\newpage

\bibliography{bibliography}

\section*{End Matter}\label{end}

{\it Appendix A: Numerical construction.---} \label{subend:numerics}
To construct the black-hole solutions discussed
in this Letter, we begin by considering
the following ansatz
\begin{align}\label{eq:ansatz}
ds^2&=\frac{1}{z^2}\left[-g(z)dt^2+\frac{dz^2}{f(z)}+d\vec{x}^{\,2}\right],\quad A_\mu=0\,,\nonumber\\
&\phi=(1-\eta)\psi(z)\,,\qquad\bar{\phi}=(1+\eta)\psi(z)\,.
\end{align}
Introducing $\kappa^2=\eta^2-1$ and
$\mathcal V(\psi)=V(-\kappa^2\psi^2)$,
with $V$ defined in Eq.~\eqref{eq:AdSpotential},
the equations of motion reduce to
\begin{align}\label{eq:app_eoms}
\partial_z f
&=\frac{3(f-1)}{z}
-\frac{\kappa^2 zf}{2}(\partial_z\psi)^2
+\frac{\mathcal V}{2z},
\nonumber\\
\partial_z^2\psi
&+\frac12\left(
\frac{\partial_z f}{f}
+\frac{\partial_z g}{g}
-\frac4z
\right)\partial_z\psi
=-\frac{\mathcal V'}{2\kappa^2 fz^2},
\nonumber\\
\frac{\partial_z g}{g}
&=\frac{\partial_z f}{f}
+\kappa^2 z(\partial_z\psi)^2\,,
\end{align}
where $\mathcal V'=d\mathcal V/d\psi$. Using the scaling symmetry of the background,
we fix the event horizon at $z_h=1$.
Regularity at the horizon requires
\begin{align}
    f&=U_1(1-z)+\cdots\,,\qquad g=G_1(1-z)+\cdots\,,\nonumber\\
    \psi&=\psi_h-\Psi_1(1-z)+\cdots\,,
    \label{eq:nearhorizonExp}
\end{align}
where 
\begin{equation}
U_1=3-\frac{\mathcal V(\psi_h)}{2},
\qquad
\Psi_1=\frac{\mathcal V'(\psi_h)}
{2\kappa^2 U_1}\,.
\label{eq:app_horizon}
\end{equation}
We focus on the NEC-violating regime,
$\kappa^2>0$, and set $m^2=-2$.
For fixed $\kappa$ and $v$, we solve
Eq.~\eqref{eq:app_eoms} numerically by
shooting from the event horizon toward the
AdS boundary at $z=0$, imposing $g(0)=1$
and $\psi'(0)=M$.

\begin{figure}[t]
    \centering
    \includegraphics[width=0.99\linewidth]
    {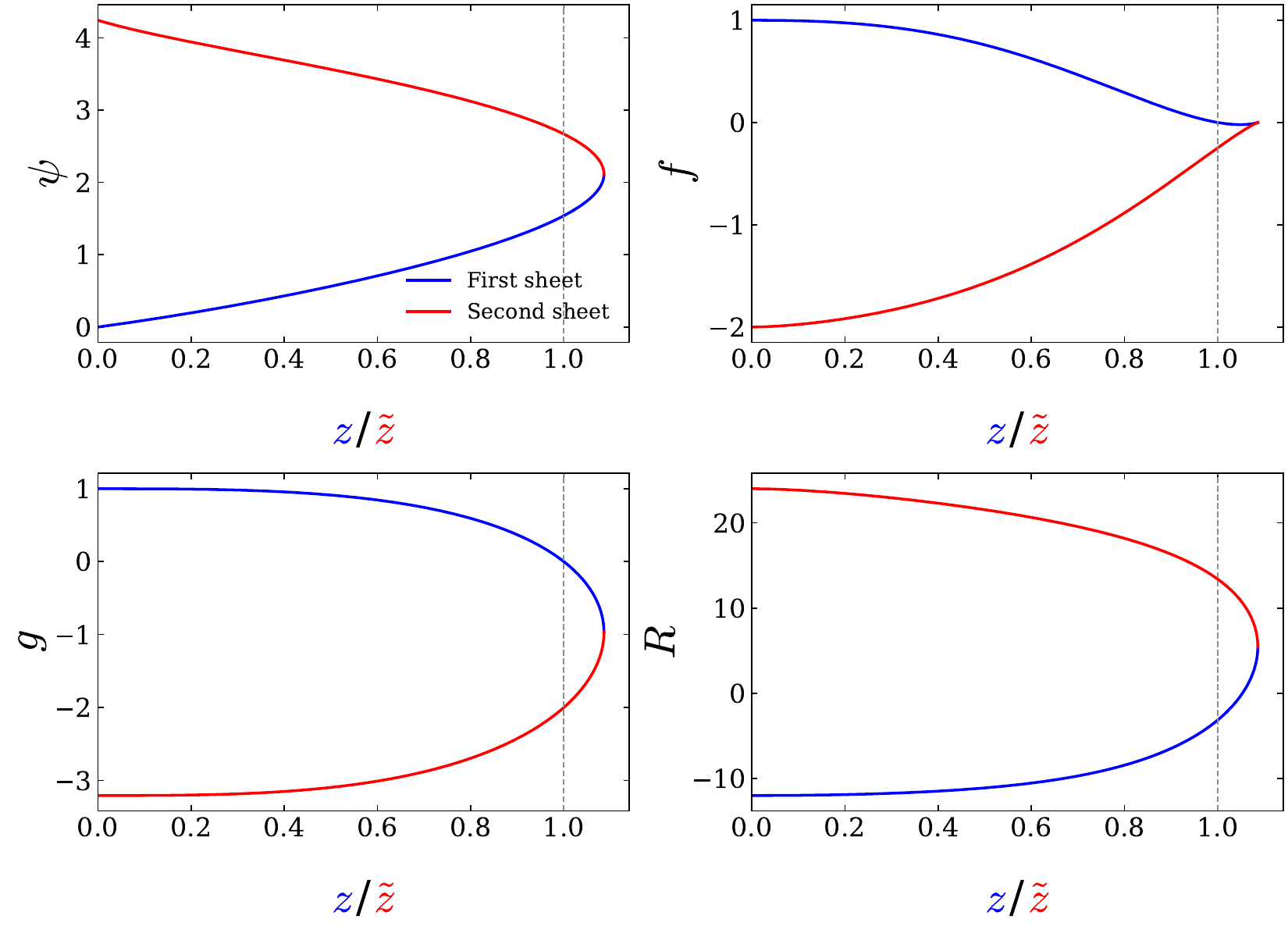}\par
\vspace{-2mm}
\caption{Representative black-hole solution
with a dS interior for $v=1/18$ and
$\sqrt{\eta^2-1}\,M/T=50/9$.
Blue curves trace the first radial patch
from the AdS boundary through the event
horizon (dotted line) to $z_\sigma$,
while red curves follow the second sheet
toward $\tilde z=0$.
Although $f$ vanishes at $z_\sigma$,
the geometry remains regular in the
signed radial coordinate $r$, with the
Ricci scalar $R$ (and other curvature
invariants) staying finite.
The fields approach constant values
at the dS endpoint, in contrast
to the divergent Little Rip evolution
of Ref.~\cite{Arean:2026jie}.}
    \label{fig:solution_full_range}
\end{figure}

To construct the interior, we continue the
numerical integration beyond the event horizon.
As in Ref.~\cite{Arean:2026jie}, solutions
with dS interiors become non-analytic in the
coordinate $z$ at a finite $z=z_\sigma$.
This coordinate singularity is resolved by
introducing the coordinate
$r=\sqrt{z_\sigma-z}$, which allows a smooth
continuation through $r=0$ to $r<0$.
The latter defines a second radial patch/sheet,
with $\tilde z=z_\sigma-r^2$
(see Ref.~\cite{Arean:2026jie} for details).
Along this second sheet, $\tilde z$ decreases
from $z_\sigma$ to $0$, where the geometry
approaches an asymptotic dS solution,
\begin{equation}
f\to-H^2,\qquad
g\to-c_0^2,\qquad
\psi\to\psi_*=-\frac{i\rho_*}{\kappa},
\end{equation}
with $\rho_*^2=-1/v$ as defined below
Eq.~\eqref{eq:stuckelberg}.
Using the coordinate transformation
$\tilde z=-H\tau$, we recover
the asymptotic dS metric~\eqref{eq:dsmetric}.
A representative numerical solution, including both
radial sheets, is shown in
Fig.~\ref{fig:solution_full_range}.

\begin{figure}[t]
    \centering
    \includegraphics[width=0.99\linewidth]
    {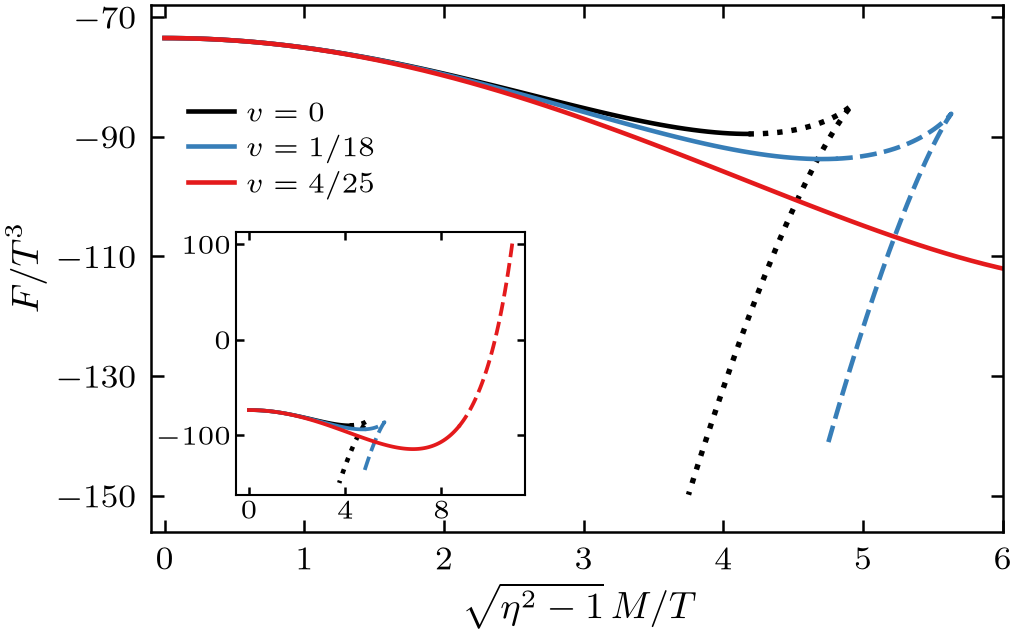}\par
\vspace{-2mm}
\caption{Dimensionless free-energy density
$F/T^3$ as a function of
$\sqrt{\eta^2-1}\,M/T$, for $v=0$, $1/18$ and $4/25$.
The solid segments correspond to black holes
with a Kasner interior, the dotted segments to a
Little Rip interior, and the dashed segments
to a de Sitter interior.}
    \label{fig:free_energy}
\end{figure}

As in the Little Rip case~\cite{Arean:2026jie},
we find multiple branches of black-hole solutions
at fixed non-Hermitian deformation. To identify
the thermodynamically preferred branch, we compare
their free energies, shown in
Fig.~\ref{fig:free_energy} for representative
values of the quartic coupling.

\vspace{1.5mm}
{\it Appendix B: Holographic renormalization for AdS.---} \label{subend:AdSHoloRenrom}
To compute the AdS boundary one-point functions,
we renormalize the bulk on-shell action~\eqref{eq:action} following
standard holographic renormalization
\cite{Skenderis:2002wp}.
For $m^2=-2$, the required boundary terms,
evaluated at $z=\epsilon$, are
\begin{equation}
S_{\rm ct}=\int d^3x\sqrt{-\gamma}
\left[2K-(4+\phi\bar{\phi})\right],
\end{equation}
where $\gamma_{\mu\nu}$ is the induced metric
at the cutoff surface and $K$ is the trace
of its extrinsic curvature.

Near the AdS boundary, the solutions of
Eq.~\eqref{eq:app_eoms} admit the expansions
\begin{align}
f&\approx 1-\frac{\kappa^2M^2}{2}z^2
-\left(g_3+\frac43\kappa^2M\psi_{\rm v}\right)z^3\,,
\nonumber\\
g&\approx 1-g_3z^3\,,
\nonumber\\
\psi&\approx Mz+\psi_{\rm v}z^2\,.
\label{eq:AdSexpansion}
\end{align}
Varying the renormalized on-shell action
with respect to the boundary sources yields
\begin{align}\label{eq:1ptfuncs_AdS}
\expval{\mathcal O}
&=(1-\eta)\psi_{\rm v}\,,
\qquad
\expval{\bar{\mathcal O}}=(1+\eta)\psi_{\rm v}\,,
\nonumber\\
\expval{T_{tt}}
&=2g_3+\frac23\kappa^2M\psi_{\rm v}\,,
\nonumber\\
\expval{T_{11}}
&=\expval{T_{22}}
=g_3-\frac23\kappa^2M\psi_{\rm v}\,.
\end{align}
In particular, the stress-tensor components
satisfy
\begin{equation}
\label{eq:wardid}
-\left(
\expval{T^t{}_t}-\expval{T^1{}_1}
\right)=3g_3=sT,
\end{equation}
providing the AdS contribution to the
identity~\eqref{eq:identity}.

The renormalized Euclidean on-shell action also
determines the free-energy density plotted in
Fig.~\ref{fig:free_energy},
\begin{equation}
F=\frac{S_{E,\rm ren}}{\beta V_2}
=-g_3+\frac23\kappa^2M\psi_{\rm v},
\label{eq:AdSfreeenergy}
\end{equation}
where we used $sT=3g_3$ from
Eq.~\eqref{eq:wardid}.

\vspace{1.5mm}

{\it Appendix C: Holographic renormalization for dS.---}
\label{subend:dSHoloRenrom}
To compute the one-point functions of QFT$^+$,
we renormalize the bulk on-shell
action~\eqref{eq:dSaction} by adding boundary
terms at $\mathcal I^+$. For $m^2=-2$ and $v=1/18$ ($H^2=2$),
the required boundary terms, evaluated
at $\tau=-\epsilon$, are
\begin{equation}
S_{\rm ct}=-\int d^3x\sqrt h\,
[2K-H(4+\sigma^2+2v\rho_*\sigma^3\log(-\mu\tau))].
\end{equation}
where $h_{\mu\nu}$ is the induced metric at
the cutoff surface, $K$ is the trace of its
extrinsic curvature, and $\mu$ is an arbitrary
renormalization scale.

Using the coordinate
$\tilde z=-H\tau$, the asymptotic solutions
near $\mathcal I^+$ admit the following expansions:
\begin{align}
f&\approx-H^2+\kappa^2\delta\psi_1^2\tilde z^2
+\tilde z^3\left[
\delta f_3-\frac{4H\kappa^3}{3}
\delta\psi_1^3\log\tilde z
\right],
\nonumber\\
g&\approx-c_0^2+\frac{c_0^2}{2}\tilde z^3
\left[
\delta f_3-\frac83\kappa^2
\delta\psi_1\delta\psi_2
+\frac{2H}{9}\kappa^3\delta\psi_1^3
\right],
\nonumber\\
\psi&\approx\psi_*+\delta\psi_1\tilde z
+\tilde z^2\left[
\delta\psi_2-\frac{\kappa}{H}
\delta\psi_1^2\log\tilde z
\right].
\label{eq:dSexpansion}
\end{align}
As expected for our parameter choices,
the scalar exhibits two characteristic
falloffs, proportional to $\tilde z$ and
$\tilde z^2$. Nonlinear interactions generate
the logarithmic corrections.

Following the dS/CFT prescription
\eqref{eq:ZQFTplus}, the Euclidean one-point
functions are defined by functional
derivatives with respect to the boundary
sources,
\begin{align}
\expval{\mathcal O_\sigma}
&=\frac{1}{\sqrt{h_{(b)}}}
\frac{\delta\log\mathcal Z_{{\rm QFT}^+}}
{\delta\sigma_s}
=\frac{i}{c_0}
\frac{\delta S_{\rm ren}}{\delta\sigma_s},
\nonumber\\
\expval{\mathcal T_{\mu\nu}}
&=-\frac{2}{\sqrt{h_{(b)}}}
\frac{\delta\log\mathcal Z_{{\rm QFT}^+}}
{\delta h_{(b)}^{\mu\nu}}
=-\frac{2i}{c_0}
\frac{\delta S_{\rm ren}}
{\delta h_{(b)}^{\mu\nu}},
\label{eq:dS1ptprescription}
\end{align}
where $h_{(b)\mu\nu}=\mathrm{diag}(c_0^2,1,1)$
is the Euclidean boundary metric and
$\sigma_s=-i\kappa H\delta\psi_1$ is the
source of the scalar operator.
Evaluating these variations in terms of
the asymptotic data yields
\begin{align}\label{eq:1ptfuncs_dS}
\expval{\mathcal O_\sigma}
={}&-2\kappa\,\delta\psi_2
+H\kappa^2(\delta\psi_1)^2
\left(1-\log\frac{\mu}{H}\right),
\nonumber\\
\expval{\mathcal T_{tt}}
={}&2ic_0^2\left[
\frac{\delta f_3}{H}
-H\kappa^2\delta\psi_1\delta\psi_2
+\frac{\kappa^3(\delta\psi_1)^3}{3}
\log\frac{\mu}{H}
\right],
\nonumber\\
\expval{\mathcal T_{11}}
={}&\expval{\mathcal T_{22}}
\nonumber\\
={}&2i\left[
-\frac{\delta f_3}{2H}
+H\kappa^2\delta\psi_1\delta\psi_2
\right.
\left.
-\frac{\kappa^3(\delta\psi_1)^3}{3}
\left(1-\log\frac{\mu}{H}\right)
\right].
\end{align}
For real bulk solutions, the Euclidean
stress tensor is purely imaginary, as
expected from the wavefunction prescription.
Its components reproduce the dS side
of the stress-tensor identity~\eqref{eq:identity}.

Finally, these one-point functions satisfy
the trace Ward identity
\begin{equation}
\expval{\mathcal T^\mu{}_\mu}
-\sigma_s\expval{\mathcal O_\sigma}
=\frac{2\sigma_s^3}{3H^3}\,,
\label{eq:dSWard}
\end{equation}
where the right-hand side represents the local
trace anomaly arising from the logarithmic
counterterm.

\newpage
\onecolumngrid 
\clearpage

% Full-width footnotes in Supplemental Material
\makeatletter
\let\set@footnotewidth\set@footnotewidth@one
\let\compose@footnotes\compose@footnotes@one
\onecolumn@grid@setup
\makeatother

\section*{Supplemental Material}
\setcounter{section}{0}

This Supplemental Material provides technical details
for the three holographic probes studied in the Letter.
Section~\ref{sec:SM_geodesics} extracts asymptotic dS data
from two-sided correlators; Sec.~\ref{sec:SM_complexity}
analyzes the distinct regimes of holographic complexity;
and Sec.~\ref{sec:SM_chaos} investigates black-hole and
cosmological-horizon scrambling and its observer interpretation.

\section{Two-sided correlators and asymptotic \lowercase{d}S data}
\label{sec:SM_geodesics}

We probe the emergent dS region through analytically
continued two-sided AdS correlators. Using dS asymptotics,
we show how nonanalytic terms in the geodesic length
encode renormalized data at $\mathcal I^+$.

\subsection{Two-sided geodesics and the near-null limit}

For a heavy probe operator $\mathcal O_H$ with $\Delta_H\gg1$,
the thermofield-double two-sided correlator is approximated by a
spacelike geodesic joining the AdS boundaries
\cite{Fidkowski:2003nf,Frenkel:2020ysx,Arean:2026jie}, 
\begin{equation}
G_{LR}(t)=\expval{\mathcal O_H(t-i\beta/2)\mathcal O_H(-t)}
\sim e^{-\Delta_H L_{\rm ren}(t)}.
\label{eq:SM_corr}
\end{equation}
Here $\Delta_H$ is distinct from the dS scaling dimensions
$\Delta_\pm$ of the background scalar. 
We use the metric ansatz
\eqref{eq:ansatz},
\begin{equation}
ds^2=\frac{1}{z^2}\left[-g(z)dt^2+\frac{dz^2}{f(z)}
+d\vec x^{\,2}\right],
\label{eq:SM_metric}
\end{equation}
and parameterize a geodesic at fixed $\vec x$ by its proper length
$L$. The conserved energy $E=g\dot t/z^2$ and normalization
$g_{\mu\nu}\dot x^\mu\dot x^\nu=1$ yield
\begin{equation}
\begin{aligned}
\frac{dt}{dz}&=\frac{Ez}{\sqrt{fg(E^2z^2+g)}},
&\qquad \frac{dL}{dz}&=\frac1z\sqrt{\frac{g}{f(E^2z^2+g)}}.
\end{aligned}
\label{eq:SM_geo_time}
\end{equation}
These equations describe the trajectory of a geodesic that crosses the event horizon and extends into the
second radial patch $\tilde z$. The turning points is determined by
$E^2z_*^2+g(z_*)=0$ (or $\tilde z_*$ on the second sheet).
As $E\to\infty$, the turning point approaches
$\mathcal I^+$, while the boundary time approaches
and 
$\Delta t\equiv t-t_{\rm null}\to0$.
The renormalized geodesic length entering
Eq.~\eqref{eq:SM_corr} is defined by
$L_{\rm ren}=\lim_{\epsilon\to0}
[L(\epsilon)+2\log\epsilon]$.

\subsection{Nonanalytic geodesic terms and one-point functions}

On the second radial sheet, $\tilde z=-H\tau\to0^+$ at
$\mathcal I^+$. For $m^2=-2$ and $v=1/18$, 
the asymptotic
dS behavior is given by Eq.~\eqref{eq:dSexpansion}.
The turning-point condition gives 
$\tilde z_*=c_0/E+O(E^{-4})$, explicitly showing that
near-null geodesics probe the asymptotic dS region. 

Performing a large-$E$ expansion we find that $L_{\rm ren}$ can be decomposed in three pieces: $L_{\rm ren}=L_{\rm AdS}+L_{\rm dS}+L_{\rm reg}$; where $L_{\rm reg}$ contains only analytic terms in the expansion while $L_{\rm AdS}$ and $L_{\rm dS}$ contain the non-analytic contributions arising from the AdS boundary and $\mathcal{I}^+$, respectively. 
In terms of $\Delta t$, 
the AdS contribution is given by
\begin{equation}
L_{\rm AdS}=2\log(2\Delta t)
-\expval{T_{11}}(\Delta t)^3\log\Delta t+\cdots,
\label{eq:SM_LAdS}
\end{equation}
where $\expval{T_{11}}$ is $x_1x_1$ component of the energy momentum tensor of the QFT living at the AdS boundary, given in equation \eqref{eq:1ptfuncs_AdS}.
The dS contribution is 
\begin{equation}
L_{\rm dS}=\frac{\pi}{H}
+(c_0\Delta t)^3\big[\mathcal A_{\log}\log(c_0\Delta t)+C_{\log^2}\log^2(c_0\Delta t)\big]
+\cdots,
\label{eq:SM_LdS}
\end{equation}
where the coefficients are
\begin{equation}\label{eq:SM_coefsLdS}
\mathcal A_{\log}=\frac{i}{2}\left[\expval{\mathcal T_{11}}+i v\rho_*\sigma_s^3\log\frac{\mu}{2H}\right],\qquad
C_{\log^2}=-\frac{v\rho_*\sigma_s^3}{4}.
\end{equation}
Here $\expval{\mathcal T_{11}}$ is the renormalized
stress tensor of QFT$^+$, given in Eq.~\eqref{eq:1ptfuncs_AdS} and $\sigma_s$ is the source of the scalar operator $\mathcal{O}_\sigma$.
Notably, the constant term $\pi/H$
is universal and retains the same form for any $v$.

Although the $\mathcal A_{\log}$ contribution mixes with
the $(\Delta t)^3\log\Delta t$ term in
Eq.~\eqref{eq:SM_LAdS}, the dS contribution can be
isolated by an observer in the AdS boundary QFT,
since the AdS term is entirely determined by its
own one-point functions.
Therefore, the nonanalytic coefficients in Eq.~\eqref{eq:SM_LdS}
can be determined solely from the two-sided correlator and can then be used reconstruct the scalar source
and stress tensor of QFT$^+$. Inverting
Eq.~\eqref{eq:SM_coefsLdS}, we obtain
\begin{align}
\sigma_s^3&=-\frac{4 C_{\log^2}}{v\rho_*},
\nonumber\\
\expval{\mathcal T_{11}}
&=-2i\left[\mathcal A_{\log}
-2C_{\log^2}\log\frac{\mu}{2H}\right].
\label{eq:SM_reconstruction}
\end{align}
Thus, the nonanalytic structure of AdS two-sided correlators 
encodes QFT$^+$ data: $C_{\log^2}$
determines the cube of the scalar source, while
$\mathcal A_{\log}$ fixes the $x_1x_1$ component of the stress tensor after
subtracting its local scalar contribution.

\section{Constant mean curvature slices and holographic complexity}
\label{sec:SM_complexity}

We analyze AdS-anchored CMC
slices in the CAny  framework
\cite{Belin:2021bga}. A positive minimum of the interior
mean curvature separates linear late-time growth from
a finite-time approach to $\mathcal I^+$, as in
dS static-patch holography.

\subsection{CMC prescription and stationary slices}

\begin{figure}[t]
    \centering
    \includegraphics[clip, trim=0 8mm 0 5mm, width=0.4\columnwidth]{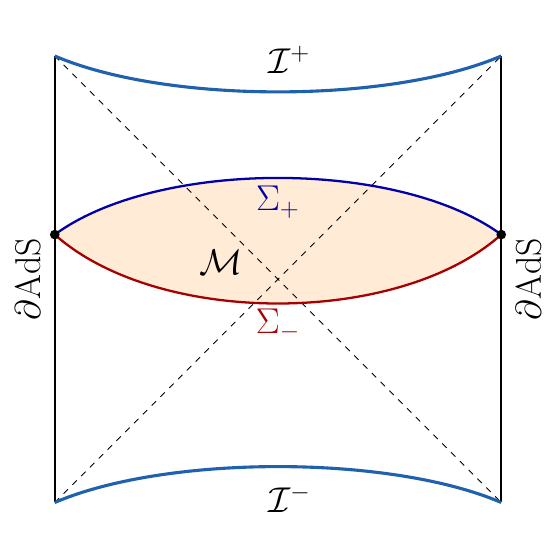}
    \caption{CMC construction with $\alpha_+=\alpha_-$.
    The shaded region $\mathcal M$ lies between AdS-anchored
    slices $\Sigma_+$ (blue) and $\Sigma_-$ (red).
    Dashed diagonals mark the event/white-hole horizons. The upper and lower boundaries
    are $\mathcal I^\pm$.}
    \label{fig:SM_CMC_region}
\end{figure}

The CAny prescription determines a bulk region $\mathcal M$
bounded by spacelike hypersurfaces $\Sigma_\pm$,
as shown in Fig.~\ref{fig:SM_CMC_region}.
These surfaces are obtained by extremizing
\begin{equation}
I_{\rm CMC}=\frac{1}{G_N}\left[
\sum_{\epsilon=\pm}\alpha_\epsilon
\int_{\Sigma_\epsilon}\!d^3\sigma\sqrt{h}
+\alpha_B\int_{\mathcal M}\!d^4x\sqrt{-g_{\rm bulk}}
\right].
\label{eq:SM_CMCaction}
\end{equation}
With future-directed unit normals and $K=\nabla_\mu n^\mu$,
the embedding variation imposes
\begin{equation}
K\big|_{\Sigma_\epsilon}=-\epsilon\frac{\alpha_B}{\alpha_\epsilon}
\equiv K_{\Sigma_\epsilon}.
\label{eq:SM_CMCcondition}
\end{equation}
Hereafter, $K_\Sigma$ denotes the prescribed
curvature, with the branch label suppressed.
The corresponding observable is
\begin{equation}
C^\epsilon=\frac{1}{G_N}\int_{\Sigma_\epsilon}
\!d^3\sigma\sqrt{h}\,\mathcal F,
\label{eq:SM_Cfunctional}
\end{equation}
Here $\mathcal F$ is a scalar geometric functional; the CV
choice is $\mathcal F=1$, $K_\Sigma=0$. We consider
reflection-symmetric two-sided slices with total boundary time
$t_b=t_L+t_R$ and $t_L=t_R=t_b/2$.

Using the ansatz~\eqref{eq:ansatz} with $z_h=1$, define
$q(z)=\sqrt{g(z)/f(z)}$ and the ingoing coordinate
$dw=dt-q(z)\,dz/g(z)$. The metric becomes
\begin{equation}
ds^2=\frac1{z^2}\left[-g(z)\,dw^2-2q(z)\,dw\,dz
+d\vec x^{\,2}\right].
\label{eq:SM_CEF}
\end{equation}
For $w=w(z)$, the volume density is
$\sqrt h=z^{-3}\sqrt{-g(z)\,w'^2-2q(z)\,w'}$, with primes denoting
$z$-derivatives. Integrating the bulk volume in
Eq.~\eqref{eq:SM_CMCaction} gives
\begin{equation}
\mathcal L_{\rm eff}=\frac{\sqrt{-g(z)\,w'^2-2q(z)\,w'}}{z^3}
+K_\Sigma b(z)w',\qquad b'(z)=\frac{q(z)}{z^4}.
\label{eq:SM_Leff}
\end{equation}
The sign of $K_\Sigma$ selects the future or past branch.
Since $w$ is cyclic, its conjugate momentum
\begin{equation}
P=\frac{\partial\mathcal L_{\rm eff}}{\partial w'}
=K_\Sigma b(z)-\frac{g(z)\,w'+q(z)}
{z^3\sqrt{-g(z)\,w'^2-2q(z)\,w'}}
\label{eq:SM_P}
\end{equation}
is conserved. 
Setting $A(P,z)=P-K_\Sigma b(z)$ yields
\begin{equation}
\dot z^2+U(P,z)=0,\qquad
U(P,z)=-\frac{g(z)}{z^6}-A(P,z)^2,
\label{eq:SM_mechanics}
\end{equation}
where the dot denotes an auxiliary affine parameter.
The surface obeys $U(P,z)\leq0$, with turning point $U(P,z_t)=0$.
The momentum also gives
\begin{equation}
\frac{dt}{dz}=-\frac{q(z)}{g(z)}\,
\frac{A(P,z)}{\sqrt{-U(P,z)}}.
\label{eq:SM_tprime}
\end{equation}
Integrating over both halves gives $t_b(P)$, with the
appropriate coordinate on each sheet. The complexity is
\begin{equation}
C^\epsilon(P)=\frac{2V_2}{G_N}
\int_{\mathcal C_P}\!dz\,
\frac{q(z)\,\mathcal F(z)}{z^6\sqrt{-U(P,z)}},
\label{eq:SM_Cintegral}
\end{equation}
Here $V_2=\int dx\,dy$ and $\mathcal C_P$ is half the
connected slice; the time-independent AdS subtraction is understood.

Late-time growth is governed by stationary
final slices, identified with double roots
of the effective potential,
\begin{equation}
U(P_f,z_f)=\partial_zU(P_f,z_f)=0,
\qquad \partial_z^2U(P_f,z_f)<0.
\label{eq:SM_double}
\end{equation}
To interpret this geometrically, introduce the constant-radius
volume density in the black-hole interior,
\begin{equation}
W(z)\equiv\frac{\sqrt{-g(z)}}{z^3},\qquad
U(P,z)=W(z)^2-A(P,z)^2.
\label{eq:SM_W}
\end{equation}
At the double root $A(P_f,z_f)=sW(z_f)$, $s=\pm1$, and
$\partial_z A(P,z)=-K_\Sigma b'(z)$ implies $W'(z_f)=-sK_\Sigma b'(z_f)$. 
The future-directed normal to a constant-$z$ slice on
the first interior sheet is $n^z=z\sqrt{-f(z)}$. Its mean
curvature profile is therefore
\begin{align}
K(z)&=\frac{1}{\sqrt{-g_{\rm bulk}}}
\partial_z\big(\sqrt{-g_{\rm bulk}}\,n^z\big)=\sqrt{-f(z)}\left[\frac{zg'(z)}{2g(z)}-3\right]
=\frac{W'(z)}{b'(z)},
\label{eq:SM_K}
\end{align}
Since $\sqrt{-g_{\rm bulk}}=q(z)/z^4=b'(z)$, the double-root
condition gives
\begin{equation}
|K(z_f)|=|K_\Sigma|.
\label{eq:SM_match}
\end{equation}
Moreover, differentiating $U(P,z)=W(z)^2-A(P,z)^2$ twice at the
double root gives
\begin{equation}
U''(P_f,z_f)=2W(z_f)b'(z_f)K'(z_f).
\label{eq:SM_Usecond}
\end{equation}
Since $W(z_f),b'(z_f)>0$, a stable final slice has $K'(z_f)<0$.
Near a nondegenerate double root, $-U(P_f,z)\propto(z-z_f)^2$,
so $t_b\sim-\gamma_f\log|z_t-z_f|$ ($\gamma_f>0$):
the slice reaches $z_f$ only as $t_b\to\infty$.
For regular $\mathcal F(z_f)$, late-time complexity is
\begin{equation}
\begin{aligned}
\lim_{t_b\to\infty}\frac{dC^\epsilon}{dt_b}
&=\frac{V_2}{G_N}\mathcal F(z_f)W(z_f),\\
C^\epsilon(t_b)&=\frac{V_2}{G_N}\mathcal F(z_f)W(z_f)\,t_b+O(1).
\end{aligned}
\label{eq:SM_rate}
\end{equation}
These equations determine the existence and late-time growth
of complexity directly from the background mean-curvature profile.

\subsection{Curvature gap and finite-time growth}

We now examine the interior profile $K(z)$,
which differs sharply from that of black holes
with Kasner singularities
\cite{Jorstad:2023kmq,Arean:2024pzo}.
Near the horizon, Eq.~\eqref{eq:nearhorizonExp}
gives $K(z)\simeq\sqrt{U_1}/(2\sqrt{z-1})
\to+\infty$, as constant-$z$ slices become null. The geometry continues through the
regular coordinate turning point $z=z_\sigma$ to the second
sheet, on which future evolution corresponds to decreasing
$\tilde z$. The normal changes orientation,
$n^{\tilde z}=-\tilde z\sqrt{-f(\tilde z)}$, giving
\begin{equation}
K(\tilde z)=-\sqrt{-f(\tilde z)}
\left[\frac{\tilde z\,g'(\tilde z)}{2g(\tilde z)}-3\right].
\label{eq:SM_Ksecond}
\end{equation}
As $\tilde z\to0$, $f(\tilde z)\to-H^2$, $g(\tilde z)\to-c_0^2$,
and $K(\tilde z)\to3H$, the mean curvature of future-directed
constant-proper-time slices in expanding dS. This finite,
positive limit contrasts with the Kasner behavior.

For the lower-free-energy solutions with emergent dS interiors,
the numerical profiles remain positive throughout the smooth
continuation and attain a nonzero minimum near the radial
turning point,
\begin{equation}
K_{\rm c}\equiv\min_{\rm interior}K>0,
\qquad 0<K_{\rm c}<3H.
\label{eq:SM_Kc}
\end{equation}
Combining Eq.~\eqref{eq:SM_Kc} with
Eq.~\eqref{eq:SM_match} gives two regimes,
\begin{equation}
\begin{array}{ll}
|K_\Sigma|>K_{\rm c}: & \text{stationary final slice,}\\[2pt]
|K_\Sigma|<K_{\rm c}: & \text{turning point reaches }\mathcal I^+.
\end{array}
\label{eq:SM_Kgap}
\end{equation}
In the first regime, $K(z)$ decreases from infinity
at the horizon and reaches $|K_\Sigma|$ before
attaining its minimum. The CMC slices approach
this constant-radius surface at late times,
producing the linear growth in
Eq.~\eqref{eq:SM_rate}.
In the second regime, no constant-radius slice
has the required mean curvature. Instead,
the turning point moves through the smooth
bounce toward $\mathcal I^+$, reaching
$\tilde z_t\to0$ as $t_b\to t_{\rm crit}<\infty$.
At the threshold $|K_\Sigma|=K_{\rm c}$,
we have $K'(z_c)=0$ and $U''(P_c,z_c)=0$,
marking the transition between the two regimes.
Both behaviors are determined by the interior
geometry, while the slices remain anchored
at the AdS boundaries.

We now derive the finite-time complexity growth
for $|K_\Sigma|<K_{\rm c}$ from the asymptotic
dS metric on the second sheet,
\begin{equation}
ds^2\simeq\frac1{\tilde z^2}
\left[c_0^2\,dt^2+dx^2+dy^2-
\frac{d\tilde z^2}{H^2}\right].
\label{eq:SM_dSmetric_C}
\end{equation}
In this regime, $q(\tilde z)\to q_0\equiv c_0/H$ and
$b'(\tilde z)\simeq q_0/\tilde z^4$.
For the volume prescription ($K_\Sigma=0$,
$\mathcal F=1$), the effective potential
in Eq.~\eqref{eq:SM_mechanics} becomes
\begin{equation}
U(P,\tilde z)\simeq\frac{c_0^2}{\tilde z^6}-P^2,
\qquad
\tilde z_t=\left(\frac{c_0}{|P|}\right)^{1/3}.
\label{eq:SM_UdS}
\end{equation}
The limit $|P|\to\infty$ therefore pushes the turning
point to $\mathcal{I}^+$. Substituting into
Eq.~\eqref{eq:SM_Cintegral}, the dS part of the
complexity satisfies
\begin{equation}
C_{\rm dS}\simeq\frac{2V_2}{G_NH\,\tilde z_t^2}
\int_1^{\tilde z_m/\tilde z_t}\!
\frac{dy}{y^6\sqrt{1-y^{-6}}}\simeq\frac{V_2}{3G_NH}
B\!\left(\frac56,\frac12\right)\tilde z_t^{-2},
\label{eq:SM_Cdivergence}
\end{equation}
where $y=\tilde z/\tilde z_t$ and $\tilde z_m$
is a fixed matching radius, while $B$ denotes
the Euler beta function.

The boundary time can be analyzed in the same approximation.
For the CV slice, Eq.~\eqref{eq:SM_tprime} reduces, up to the
orientation of the integration contour, to
\begin{equation}
\left|\frac{dt}{d\tilde z}\right|
=\frac1{Hc_0}\,
\frac1{\sqrt{1-(\tilde z_t/\tilde z)^6}}.
\label{eq:SM_tprime_dS}
\end{equation}
Its integral has a finite limit as $\tilde z_t\to0$.
Subtracting that limit yields the leading dependence
\begin{align}
|t_{\rm crit}-t_b|&=c_t\,\tilde z_t
+o(\tilde z_t),\qquad
c_t=-\frac{1}{3Hc_0}\,B\!\left(-\frac16,\frac12\right)>0,
\label{eq:SM_time_scaling}\\
C_{\rm dS}&\propto |t_{\rm crit}-t_b|^{-2},\qquad
\left|\frac{dC_{\rm dS}}{dt_b}\right|
\propto |t_{\rm crit}-t_b|^{-3}.
\label{eq:SM_hyperfast_scaling}
\end{align}
The coefficient follows from subtracting the finite limiting
time in Eq.~\eqref{eq:SM_tprime_dS}, using $t_b=t_L+t_R$.
Both exponents depend only on the asymptotic dS geometry.

The same scaling mechanism extends beyond CV.
For a future-directed CMC slice with
$|K_\Sigma|<K_{\rm c}$, define
$k\equiv K_\Sigma/(3H)$, with
$0\leq k<K_{\rm c}/(3H)<1$.
Since $b(\tilde z)\simeq-q_0/(3\tilde z^3)$, the
asymptotic momentum and potential are
\begin{align}
P&\simeq(1-k)\frac{c_0}{\tilde z_t^3},
\nonumber\\
-U(P,\tilde z)&\simeq\frac{c_0^2}{\tilde z_t^6}\left[
\big(1-k+ky^{-3}\big)^2-y^{-6}\right],
\qquad y=\frac{\tilde z}{\tilde z_t}.
\label{eq:SM_UdS_CMC}
\end{align}
For a geometric functional with a finite nonzero dS limit
$\mathcal F\to\mathcal F_\infty$, the leading contribution is
\begin{equation}
C_{\rm dS}\simeq\frac{2V_2\mathcal F_\infty}
{G_NH\,\tilde z_t^2}\,\mathcal I(k),\qquad
\mathcal I(k)\equiv\int_1^\infty\!dy\,
\frac{1}{y^6\sqrt{(1-k+ky^{-3})^2-y^{-6}}}.
\label{eq:SM_CdS_CMC}
\end{equation}
For $0\leq k<1$, this integral is finite, with
$\mathcal I(0)=\tfrac16 B(5/6,1/2)$. The $\tilde z_t^{-2}$ divergence therefore holds
across this family of CMC slices with
$\mathcal F_\infty\neq0$:
AdS-boundary-anchored slices reproduce the finite-time
mechanism of static-patch dS hyperfast growth.

\subsection{Near-horizon universality and numerical diagnostics}

\begin{figure}[t]
    \centering
    \includegraphics[width=0.8\columnwidth]{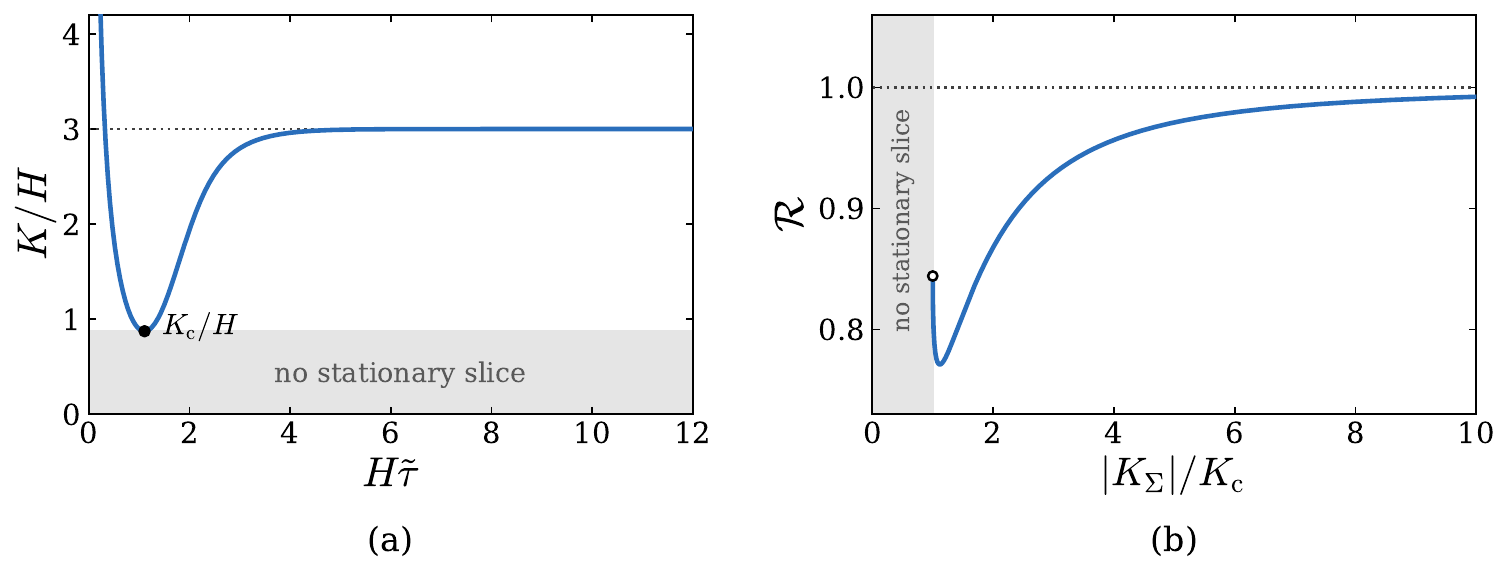}\par
\vspace{-3mm}
\caption{Mean-curvature gap and late-time complexity
growth for a dS interior ($v=1/18$,
$\sqrt{\eta^2-1}\,M/T=50/9$).
(a) $K/H$ versus interior proper time $\tilde{\tau}$ ($\tilde{\tau}=0$ at the
horizon). The dot marks $K_{\rm c}>0$; the dotted line marks
$K\to3H$. Shading indicates curvatures not attained by
constant-radius slices.
(b) Normalized rate $\mathcal R\equiv(8\pi TS)^{-1}
\lim_{t_b\to\infty}d\mathcal C^\epsilon/dt_b
=|K_\Sigma|W(z_f)/(2\pi T)$ for $\mathcal F=|K|$.
The rate is nonmonotonic and tends to $\mathcal R=1$ (dotted).
For $|K_\Sigma|<K_{\rm c}$ (shaded), no stationary final slice
exists. The open circle marks the degenerate threshold,
where Eq.~\eqref{eq:SM_rate} ceases to apply; its value is
the limit along the curve.}
    \label{fig:SM_CMC_diagnostics}
\end{figure}

We finally consider the large-curvature limit
$|K_\Sigma|\gg H$, where the stationary final
slice approaches the black-hole event horizon.
Using the near-horizon metric,
\begin{equation}
W(z)\simeq\sqrt{G_1(z-1)},\qquad
K(z)\simeq\frac{\sqrt{U_1}}{2\sqrt{z-1}}.
\label{eq:SM_KW}
\end{equation}
so their product has the universal limit
\begin{equation}
\lim_{z\to1^+}K(z)W(z)
=\frac{\sqrt{U_1G_1}}2=2\pi T.
\label{eq:SM_KW_limit}
\end{equation}
For $\mathcal F=|K|$, the late-time growth rate becomes
\begin{equation}
\lim_{|K_\Sigma|\to\infty}
\lim_{t_b\to\infty}\frac{dC^\epsilon}{dt_b}
=\frac{2\pi T V_2}{G_N}=8\pi TS,
\qquad S=\frac{V_2}{4G_N}.
\label{eq:SM_universal_rate}
\end{equation}
The limit depends only on horizon temperature and entropy;
finite-$K_\Sigma$ growth also probes the interior.
Figure~\ref{fig:SM_CMC_diagnostics} shows the mean-curvature
profile and normalized growth rate for a representative dS interior.

Figure~\ref{fig:SM_CMC_diagnostics}(a) exhibits
the positive minimum $K_{\rm c}$ and the
asymptote $K\to3H$, confirming the curvature
gap separating stationary slices from those
reaching $\mathcal I^+$ at finite time.
In panel (b), the normalized late-time growth
rate, defined for $\mathcal F=|K|$ as
$\mathcal R\equiv(8\pi TS)^{-1}
\lim_{t_b\to\infty}dC^\epsilon/dt_b$,
is nonmonotonic: it rises above the
near-horizon value, then approaches
$\mathcal R=1$ from below.
The universal limit is horizon-controlled,
while finite $|K_\Sigma|$ reveals additional
interior structure. Both regimes arise from
AdS-boundary-anchored complexity.

\section{Quantum chaos: black-hole and cosmological horizons}
\label{sec:SM_chaos}

We determine black-hole chaos parameters through
shockwaves and pole-skipping, then apply the analysis
of Ref.~\cite{Ahn:2025qat} to the asymptotic dS horizon
and explore its observer-centered interpretation.

\subsection{Black-hole scrambling and butterfly velocities}

A conventional measure of quantum scrambling is the thermal
out-of-time-order correlator
\begin{equation}
F(t,\vec x)=\frac{\langle V W(t,\vec x)V W(t,\vec x)\rangle_\beta}
{\langle VV\rangle_\beta\langle WW\rangle_\beta}.
\label{eq:SM_OTOC}
\end{equation}
In a large-$N$ holographic theory, its leading connected correction
is parameterized as
\begin{equation}
F(t,\vec x)\simeq 1-\frac{c}{N^2}
\exp\!\left[\lambda_L\left(t-\frac{|\vec x|}{v_B}\right)\right]
+\cdots,
\label{eq:SM_OTOCgrowth}
\end{equation}
for $t\ll t_*\sim(2\pi T)^{-1}\log N^2$.
Here $c$ is an $O(1)$ constant; $\lambda_L$ controls temporal growth and $v_B$
the spatial spread of boundary perturbations. In our non-Hermitian setting, the
real-time OTOC prescription in the boundary theory requires care.
We therefore extract the corresponding chaos
parameters directly from bulk gravitational
dynamics.

In AdS/CFT, an early boundary perturbation
generates a near-horizon shockwave whose
scattering with a later probe controls the
leading OTOC correction
\cite{Shenker:2013pqa,Roberts:2014isa}.
To construct this shockwave, we use the
ansatz~\eqref{eq:ansatz} and introduce
tortoise and Kruskal coordinates,
\begin{equation}
dr_*=-\frac{dz}{\sqrt{fg}},\qquad
U=-e^{2\pi T(r_*-t)},\qquad
V=e^{2\pi T(r_*+t)},
\label{eq:SM_Kruskal}
\end{equation}
where $T=\sqrt{U_1G_1}/(4\pi)$ is the Hawking
temperature.
In these coordinates the metric becomes
\begin{align}
ds^2&=2A(UV)\,dU\,dV+B(UV)d\vec x^{\,2},\nonumber\\
A&=\frac{g}{2z^2(2\pi T)^2UV},
\qquad B=\frac{1}{z^2}.
\label{eq:SM_Kruskal_metric}
\end{align}
Both $A$ and $B$ are regular at $U=0$.
A perturbation inserted at $-t_w$
produces a boosted null stress tensor,
\begin{equation}
T_{UU}^{\rm shock}\propto E e^{2\pi Tt_w}\delta(U)
\delta^{(2)}(\vec x).
\label{eq:SM_Tshock}
\end{equation}
Following Refs.~\cite{Dray:1984ha,Sfetsos:1994xa},
we describe its backreaction by a shift
$V\to V+\alpha(\vec x)$ across the horizon,
giving the perturbed metric
\begin{equation}
ds^2=2A\,dU\,dV+B\,d\vec x^{\,2}
-2A\alpha(\vec x)\delta(U)\,dU^2.
\label{eq:SM_shockmetric}
\end{equation}
The distributional construction requires $V$-independent
background fields and $T_{VV}=T_{Vi}=0$ at $U=0$.
All fields depend on $UV$, so $\partial_V\psi\propto U$
and $T_{VV}\propto(1-\eta^2)(\partial_V\psi)^2=0$
there; homogeneity also gives $T_{Vi}=0$. The construction
therefore applies despite NEC violation.

The distributional Einstein equations evaluated at the
horizon yield
\begin{equation}
\left(-\nabla_{\vec x}^2+\mu^2\right)\alpha(\vec x)
\propto E e^{2\pi Tt_w}\delta^{(2)}(\vec x),\qquad \mu^2\equiv\frac{B'(0)}{A(0)}=U_1. 
\label{eq:SM_screened}
\end{equation}
The prime in $B'(0)$ denotes an $UV$
derivative. At large separation,
\begin{equation}
\alpha(t_w,\vec x)\sim
\frac{\exp\left[2\pi T(t_w-t_*)-\mu|\vec x|\right]}
{\sqrt{|\vec x|}},
\label{eq:SM_shockprofile}
\end{equation}
up to normalization. Comparing with Eq.~\eqref{eq:SM_OTOCgrowth},
we obtain
\begin{equation}
\lambda_L=2\pi T,\qquad
v_B=\frac{2\pi T}{\mu},
\label{eq:SM_shockresult}
\end{equation}
and using $\mu^2=U_1=3-V_h/2$, we arrive at
\begin{equation}
v_B^2=\frac{(2\pi T)^2}{U_1}
=\frac{(2\pi T)^2}{3-V_h/2}.
\label{eq:SM_vB_potential}
\end{equation}
This reproduces the Letter's result.
While $\lambda_L=2\pi T$ retains its
conventional value, $v_B$ depends on the
horizon geometry and can be enhanced by
the non-Hermitian deformation.
For the Schwarzschild--AdS$_4$ black brane
with $z_h=1$, we have $U_1=G_1=3$, giving
$T=3/(4\pi)$ and $v_B=\sqrt3/2$.

As an independent check, we turn to pole-skipping,
which also determines the horizon chaos parameters
\cite{Grozdanov:2017ajz,Blake:2017ris}.
The retarded energy-density Green's function has a special point
\begin{equation}
\omega_*=i\lambda_L,\qquad
k_*^2=-\frac{\lambda_L^2}{v_B^2}.
\label{eq:SM_PSdefinition}
\end{equation}
At this point, the ingoing horizon conditions
no longer uniquely determine the bulk solution,
and the Green's function depends on the path
of approach in the $(\omega,k)$ plane.

In ingoing Eddington--Finkelstein coordinates,
$dv=dt-dz/\sqrt{fg}$, the metric becomes
\begin{equation}
ds^2=\frac{1}{z^2}\left[-g\,dv^2
-2\sqrt{\frac{g}{f}}\,dv\,dz+d\vec x^{\,2}\right].
\label{eq:SM_EF}
\end{equation}
we consider fluctuations
$\delta g_{\mu\nu}(z)e^{-i\omega v+ikx}$ and
$\delta\psi(z)e^{-i\omega v+ikx}$ in radial gauge
$\delta g_{z\mu}=0$. The sound channel contains
$\delta g_{vv}$, $\delta g_{vx}$, $\delta g_{xx}$,
$\delta g_{yy}$, and $\delta\psi$.

At the horizon, the leading $vv$ Einstein equation reduces,
after using the background equations, to
\begin{equation}
(k^2-2i\omega\sqrt{U_1/G_1})\delta g_{vv}
+(\omega-2\pi iT)
[\omega(\delta g_{xx}+\delta g_{yy})-2k\delta g_{vx}]=0.
\label{eq:SM_vv}
\end{equation}
At the special point
\begin{equation}
\omega_*=2\pi Ti,\qquad
k_*^2=-4\pi T\sqrt{\frac{U_1}{G_1}}
=-U_1=-\left(3-\frac{V_h}{2}\right)=-\mu^2,
\label{eq:SM_PSpoint}
\end{equation}
equation~\eqref{eq:SM_vv} is identically
satisfied, leaving the ingoing perturbations
less constrained. Comparing with
Eq.~\eqref{eq:SM_PSdefinition} gives precisely
Eq.~\eqref{eq:SM_shockresult}.

Shockwaves and pole-skipping thus independently
yield the same horizon chaos parameters, including
their dependence on $V_h$, providing a nontrivial
consistency check of the gravitational analysis.

To interpret the enhancement of $v_B$, recall that
for isotropic, planar, asymptotically AdS$_4$ black
branes in two-derivative Einstein gravity
with NEC-satisfying matter, Refs.~\cite{Mezei:2016wfz,Mezei:2016zxg}
establish, under their assumptions, the bound
\begin{equation}
v_B\leq\frac{\sqrt3}{2}.
\label{eq:SM_NECbound}
\end{equation}
Our real $\mathcal{PT}$-restored solutions violate the NEC,
with $\kappa^2\equiv\eta^2-1>0$. The Einstein equations
\eqref{eq:app_eoms} imply
\begin{equation}
\partial_z\log\!\left(\frac{g}{f}\right)
=\kappa^2 z\psi'(z)^2\geq0.
\label{eq:SM_chiprime}
\end{equation}
Since $g/f\to1$ at the AdS boundary, a nontrivial scalar
profile gives
$G_1/U_1=\exp[\kappa^2\int_0^1 dz\,z\psi'(z)^2]>1$.
Although NEC violation increases this ratio,
$U_1$ itself also varies. Consequently,
NEC violation allows, but does not guarantee,
superluminal butterfly velocities:
Eq.~\eqref{eq:SM_vB_potential} gives
$v_B=\sqrt{G_1}/2$, so $v_B>1$ whenever $G_1>4$.

\begin{figure}[t]
    \centering
    \includegraphics[width=0.5\columnwidth]{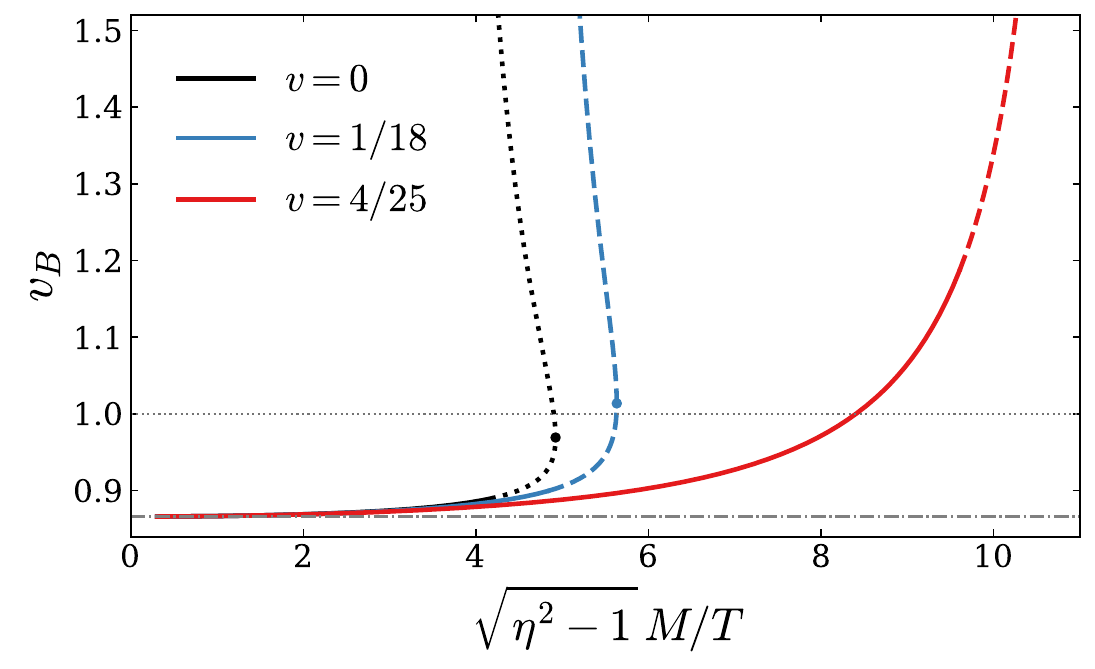}\par
\vspace{-3mm}
\caption{
Butterfly velocity $v_B$ versus
$\sqrt{\eta^2-1}\,M/T$ for $v=0$, $1/18$,
and $4/25$, using the same conventions
as Fig.~\ref{fig:free_energy}.
Solid, dotted, and dashed curves correspond
to Kasner, Little Rip, and dS interiors,
respectively.
Filled dots mark the turning points where
two solution branches meet.
Horizontal dotted and dash-dotted lines
mark $v_B=1$ and the NEC-satisfying
planar-brane bound $\sqrt{3}/2$, respectively.
The latter is saturated in the Schwarzschild--AdS$_4$
limit $\sqrt{\eta^2-1}\,M/T\to0$.
For cases with multiple branches, the lower-free-energy
branch, disconnected from the Hermitian limit,
is superluminal over nearly its entire range,
with $v_B$ increasing as the deformation decreases.
For the single-branch case, $v_B$ instead
increases with the deformation.
}
    \label{fig:butterfly_vel}
\end{figure}

Fig.~\ref{fig:butterfly_vel} illustrates the
dependence of $v_B$ on the dimensionless deformation
$\sqrt{\eta^2-1}\,M/T$, using the same parameters
and conventions as Fig.~\ref{fig:free_energy}.
For $v=0$ and $v=1/18$, two branches with
different free energies meet at a turning point.
The lower-free-energy branch, disconnected
from the Hermitian limit, exhibits $v_B>1$
over nearly its entire range, with $v_B$
increasing as the dimensionless deformation decreases.
By contrast, for $v=4/25$, the solutions form
a single branch connected to the Hermitian limit,
along which $v_B$ increases with the deformation,
eventually becoming superluminal.
The turning points do not generally coincide
with the threshold $v_B=1$.

Since $v_B$ depends only on event-horizon
data, its enhancement does not uniquely
characterize the interior geometry.
Indeed, superluminal values occur for
solutions with Kasner, Little Rip, and dS
interiors alike. Notably, superluminal
$v_B$ also occurs for small SdS black holes
without NEC violation \cite{Ahn:2025qat},
where it was conjectured to reflect nonlocality
in the putative holographic dual.
Similar enhancements arise in top-down
holographic models with nonlocal boundary
theories, including noncommutative and
dipole-deformed super Yang--Mills
\cite{Fischler:2018kwt,Eccles:2021zum}.
In our case, however, the boundary theory is
local but non-Hermitian. This raises a natural
question: do these enhancements of $v_B$ reflect
a common mechanism, or fundamentally different
microscopic dynamics?

\subsection{Time-dependent interior and the asymptotic static patch}
\label{sec:SM_cosmological_chaos}

\begin{figure}[t]
    \centering
    \includegraphics[clip, trim=0 8mm 0 5mm, width=0.4\columnwidth]{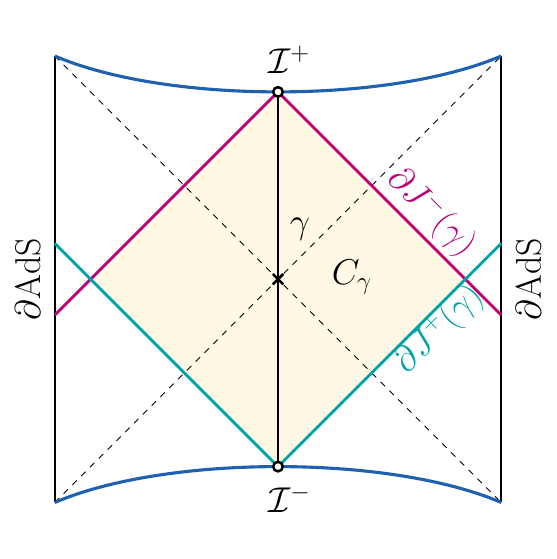}
    \caption{Radial causal section of the emergent dS
interior. Vertical lines are the timelike AdS
boundaries $\partial\mathrm{AdS}$; blue curves mark
$\mathcal I^\pm_{\rm dS}$. Dark dashed diagonals
show the black-/white-hole horizons.
A representative observer $\gamma$ determines future
(magenta) and past (teal) horizons,
$\partial J^-(\gamma)$ and $\partial J^+(\gamma)$,
bounding the shaded causal patch
$\mathcal C_\gamma=J^-(\gamma)\cap J^+(\gamma)$.
At late cosmological times, the geometry approaches
pure dS, and the observer's causal patch approaches
the corresponding dS static patch.}
    \label{fig:SM_cosmological_horizons}
\end{figure}

The cosmological horizon is observer-dependent, not a second
zero of $f(z)$. Figure~\ref{fig:SM_cosmological_horizons}
shows the causal patch of a particular comoving observer
$\gamma$, given by
$\mathcal C_\gamma=J^-(\gamma)\cap J^+(\gamma)$
and bounded by $\partial J^-(\gamma)$ and
$\partial J^+(\gamma)$. These cosmological
horizons are distinct from the global AdS
black-/white-hole horizons (dashed diagonals).\footnote{
In the degenerate limit where $\gamma$ ends at
the intersection of $\mathcal I^+$ and
$\partial\mathrm{AdS}$, a cosmological horizon
can coincide with a black-hole horizon in the
Penrose diagram. This limiting coincidence, however, does not imply an
identification of their generators or temperatures.}

On a homogeneous slice behind the event horizon, the
metric can be written in Bianchi I form,
\begin{equation}
ds^2=-d\tilde{\tau}^2+a_\parallel(\tilde{\tau})^2dt^2
+a_\perp(\tilde{\tau})^2(dx^2+dy^2),
\label{eq:SM_BianchiI}
\end{equation}
with $a_\parallel=\sqrt{-g}/z$ and $a_\perp=1/z$
(replacing $z$ by $\tilde z$ on the second sheet).
The directional rates $H_\parallel=\partial_{\tilde{\tau}}\log a_\parallel$
and $H_\perp=\partial_{\tilde{\tau}}\log a_\perp$ differ during the
intermediate cosmology, so there is no exact static-patch
Killing horizon or stationary Gibbons--Hawking temperature.

At late proper time, $f\to-H^2$, $g\to-c_0^2$
on the second radial sheet, with
$H_\parallel,H_\perp\to H$, giving
\begin{align}
ds^2&\simeq-d\tilde{\tau}^2+e^{2H\tilde{\tau}}
\left[c_0^2dt^2+dx^2+dy^2\right].
\label{eq:SM_dS_FRW}
\end{align}
Defining $\vec X=(c_0t,x,y)$ and the physical radius
$R=e^{H\tilde{\tau}}|\vec X|$, the coordinate transformation
\begin{equation}
T_s=\tilde{\tau}-\frac{1}{2H}\log(1-H^2R^2)
\label{eq:SM_static_time}
\end{equation}
recovers the pure-dS static patch,
\begin{equation}
ds^2=-(1-H^2R^2)dT_s^2
+\frac{dR^2}{1-H^2R^2}+R^2d\Omega_2^2.
\label{eq:SM_dS_static}
\end{equation}
In the asymptotic dS limit, the future and past
cosmological horizons shown in the causal diagram
become the two branches of the static-patch Killing
horizon, located at $R_c=H^{-1}$, with temperature
$T_c=H/(2\pi)$ and surface gravity $\kappa_c=H$.
Away from this limit, the anisotropic evolution
breaks the static-patch isometry.

\subsection{Cosmological scrambling and its observer interpretation}
\label{sec:SM_chaos_interpretation}

Having identified the asymptotic static patch of our
emergent cosmology, we now investigate its scrambling
properties. To this end, we apply the shockwave and
pole-skipping analysis of Ref.~\cite{Ahn:2025qat}
to the late-time dS limit of our solutions.
Since the full interior is time-dependent, these
calculations characterize the asymptotic cosmological
horizon, rather than its finite-time evolution.
The asymptotic expansion~\eqref{eq:dSexpansion}
shows that deviations from pure dS decay at
late cosmological times, with
$f+H^2=O(\tilde z^2)$ and
$g+c_0^2=O(\tilde z^3)$. 
Determining the corresponding corrections to
horizon scrambling requires a time-dependent
analysis beyond the scope of this work.

A key distinction between black-hole and
cosmological-horizon scrambling in our setup
lies in their respective generators of time
evolution. The black-hole Killing generator
$\partial_t$ governs AdS boundary time,
whereas the asymptotic cosmological horizon
is generated by
\begin{equation}
\partial_{T_s}=\partial_{\tilde{\tau}}
-HX^i\partial_{X^i}.
\label{eq:SM_static_generator}
\end{equation}
The two notions of time evolution are distinct,
suggesting that the cosmological chaos parameters
may characterize observer-centered static-patch
dynamics rather than ordinary AdS boundary OTOCs.

In the pure-dS limit, an early perturbation
generates a boosted shockwave
near the cosmological horizon. For a perturbation
inserted at static-patch time $-t_w$, its
near-horizon momentum scales as
$p\propto e^{2\pi T_c t_w}=e^{Ht_w}$.
Unlike pure dS or SdS, our geometry possesses
timelike AdS boundaries, where Lorentzian operator
insertions can generate bulk perturbations
propagating into the cosmological interior.
This offers a potential route to preparing
cosmological shockwaves from the
AdS boundary theory.
In four dimensions, the shockwave equation
of Ref.~\cite{Ahn:2025qat} becomes
\begin{equation}
\left(\nabla_{S^2}^2+2\right)\alpha(t_w,\Omega)
=\mathcal J(t_w,\Omega),
\label{eq:SM_dSshock}
\end{equation}
where the source $\mathcal J$ includes the
boost factor $e^{Ht_w}$ and is localized
on the cosmological horizon.
Unlike the screened Poisson equation at a
black-hole horizon, the positive sign of the
transverse term produces an oscillatory
spatial response. In a local angular approximation, the shockwave
profile takes the schematic form
\begin{equation}
\alpha(t_w,\theta)\sim e^{Ht_w}
\left[c_1\cos(\sqrt2\,\theta)
+c_2\sin(\sqrt2\,\theta)\right].
\label{eq:SM_dSprofile}
\end{equation}
The oscillatory dependence contrasts
with the exponential decay characteristic of
black-hole-horizon shockwaves.

The pure-dS analogue of the chaos parameters follows by
formally comparing the angular oscillation with the
conventional butterfly profile:
\begin{equation}
\lambda_{L,c}=2\pi T_c=H,\qquad
v_{B,c}^{(\theta)\,2}=-\frac{H^2}{2}.
\label{eq:SM_cosmo_angular}
\end{equation}
With physical horizon distance $\ell_\perp=\theta/H$,
this becomes $v_{B,c}^{(\mathrm{phys})\,2}=-1/2$, or
$v_{B,c}^{(\mathrm{phys})}=\pm i/\sqrt2$, with the imaginary velocity parametrizing an oscillatory response.

The gravitational sound-channel pole-skipping
analysis in pure dS independently yields
\cite{Ahn:2025qat}
\begin{equation}
\omega_{*,c}=iH,\qquad k_{S,*}^2=2.
\label{eq:SM_dSPS}
\end{equation}
Here $\omega_{*,c}$ is conjugate to static-patch
time $T_s$, while $k_{S,*}^2$ is the angular
Laplacian eigenvalue. These values reproduce
the chaos parameters obtained from the
cosmological shockwave analysis.

To interpret these cosmological chaos parameters,
we turn to the observer correlators of
Ref.~\cite{Harlow:2026pwe},
which exhibit \emph{anti-scrambling}: gravitational
backreaction can cause a probe to reach an observer
earlier, rather than delaying it as in black-hole
scrambling. In their localized dS$_3$ calculation,
the shift in arrival time is
\begin{equation}
\delta t_{\rm arr}(\phi)
=-\log[1-h(\phi)]\simeq h(\phi),
\label{eq:SM_time_advance}
\end{equation}
where $h(\phi)$ is the shockwave-induced null shift.
For light observers, recoil produces an
angle-dependent shift whose sign depends on
the direction of the incoming probe.
For sufficiently heavy observers, the angular
dependence becomes weak and gravitational
backreaction produces a time advance.
The latter underlies the anti-scrambling behavior
of observer four-point functions
\cite{Harlow:2026pwe}.

Our imaginary butterfly velocity characterizes
the oscillatory spatial response of cosmological-horizon
shockwaves. Interestingly, the localized dS$_3$
shockwave of Ref.~\cite{Harlow:2026pwe} also
exhibits an oscillatory angular profile, whose
effect on observer arrival times depends on recoil.
This suggests a concrete way to probe imaginary
butterfly velocities through gravitationally
dressed observer OTOCs: consider directionally
localized insertions and study their dependence
on relative angular separation. In our dS$_4$
setting, this could reveal the characteristic
spatial modulation of cosmological shockwaves,
while varying the observer's mass could help
disentangle recoil from gravitational
anti-scrambling.

\end{document}